\documentclass[a4paper,english,aps,manuscript]{revtex4-1}
\usepackage[T1]{fontenc}
\usepackage[latin9]{inputenc}
\usepackage{color}
\usepackage{float}
\usepackage{amsmath}
\usepackage{graphicx}

\makeatletter

\pdfpageheight\paperheight
\pdfpagewidth\paperwidth

\providecommand{\tabularnewline}{\\}

\@ifundefined{textcolor}{}
{%
 \definecolor{BLACK}{gray}{0}
 \definecolor{WHITE}{gray}{1}
 \definecolor{RED}{rgb}{1,0,0}
 \definecolor{GREEN}{rgb}{0,1,0}
 \definecolor{BLUE}{rgb}{0,0,1}
 \definecolor{CYAN}{cmyk}{1,0,0,0}
 \definecolor{MAGENTA}{cmyk}{0,1,0,0}
 \definecolor{YELLOW}{cmyk}{0,0,1,0}
}

\makeatother

\usepackage{babel}
\begin{document}

\title{Numerical Analysis of Lamellar Gratings for Light-Trapping in Amorphous
Silicon Solar Cells}

\author{D. I. Gablinger$^{1}$ and R.H. Morf$^{1}$\\
~\\
$^{1}$Condensed Matter Theory Group\\
Paul Scherrer Institute\\
CH-5232 Villigen}
\begin{abstract}
In this paper, we calculate the material specific absorption accurately
using a modal method by determining the integral of the Poynting vector
around the boundary of a specific material. Given that the accuracy
of our method is only determined by the number of modes included,
the material specific absorption can be used as a quality measure
for the light-trapping performance. We use this method to investigate
metallic gratings and find nearly degenerate plasmons at the interface
between metal and amorphous silicon (a-Si). The plasmons cause large
undesired absorption in the metal part of a grating as used in a-Si
cells. We explore ways to alleviate the parasitic absorption in the
metal by appropriate choice of the geometry. Separating the diffraction
grating from the back reflector helps, lining silver or aluminum with
a dielectric helps as well. Gratings with depth > 60nm are preferred,
and periods > 600nm are not useful. Maximum absorption in silicon
can occur for less thick a-Si than is standard. We also find that
geometric asymmetry can improve the absorption for PV. Lastly, we
investigate the modal contributions to absorption, and find that for
the most part, the absorption can not be attributed to single modes,
but comes from the interplay of a few important modes.
\end{abstract}
\maketitle

\section{Introduction}

Our goal is to study the light-trapping properties of a large class
of geometric structures, in order to obtain a qualitative overview
and understanding how performance can be improved and what kind of
structures must be avoided which will lead to a deterioration of the
light-trapping performance. A recent and comprehensive overview of
theory and experimental studies of periodic structures for lighttrapping
can be found in\textcolor{black}{{} \citep{Mokkapati:2012wa} and upper
limits have been described in \citep{Yu:2010ct}. Model calculations
with similar goals have been performed using a calculational method
based on a Rayleigh expansion \citep{6782732,7286728}, using a Fourier
modal method in \citep{Naqavi:2011yu} and finite element (FEM) based
calculations have been carried out in \citep{Isabella:2013mz}.}

In order to perform reliable calculations of spectral properties of
metallic gratings, it is known that high spatial resolution at the
nanometer scale is required. If we wish to perform calculations of
spectra even for only one-dimensional gratings with period in the
micrometer range, the length scale will encompass a range of 100 to
1000. If discretisation is required in the x-z plane, the number of
grid points will be between $10^{4}$ and $10^{6}$.

In order to allow us to make a fairly wide ranging study of different
light-trapping structures, we confine this study to lamellar geometries,
for which modal methods are very well suited. For such gratings, there
exists the possibility to solve the Helmholtz equation analytically
in terms of trigonometric and hyperbolic functions \citep{Botten:1981kq,Botten:1981tp,Sheng:1982ys,Suratteau:1983ly}.
For this, a transcendental equation for the eigenvalues must be solved.
Once the eigenvalues are known, the eigenfunctions can be expressed
analytically and used as a basis for expanding the electromagnetic
field. 

The main problem then is to find all the important eigenvalues and
eigenfunctions, which is not an easy task for such transcendental
equations, especially since these eigenvalues are complex as soon
as we deal with absorbing materials, which is the objective of any
study of light-trapping. For this reason, we use the method previously
developed by one of the present authors \citep{Morf:1995fi}, in which
the eigenfunctions are expanded in terms of orthogonal polynomials
in each domain of constant permittivity separately, and connected
between the domains by the appropriate boundary conditions for the
electromagnetic field. If the chosen polynomial basis is large enough
and thanks to exponential convergence of eigenvalues and eigenfunctions
in the size of the polynomial basis, eigenvalues and eigenvectors
will be numerically exact, limited only by the round-off errors, i.e.
by the length of the mantissa. The eigenvalues of the Helmholtz equation
are obtained by solving an algebraic eigenvalue problem for the coefficients
of the eigenvector expressed in terms of the polynomial basis functions.
This method allows the consistent calculation of eigenvalues associated
with eigenfunctions with consistently increasing number of nodes without
any gap. Of course, the accuracy of the solution of the diffraction
problem will depend on the number of modes. But, there is no further
source of calculational inaccuracy than the truncation order $N$,
i.e. the number of modes included.

If such modal method is employed, the only discretization will be
in the periodicity direction. It can be measured by the typical distance
between adjacent zeros of the polynomial of maximal order used in
the expansion of the eigenfunctions. By optimizing the computational
algorithm, the calculation of eigenvalues and eigenvectors will dominate
the solution of the system of linear equations that expresses the
boundary conditions between adjacent grating layers and the radiation
condition for the incident field and outgoing field above and below
the grating structure. Indeed, if eigenvalues and eigenfunctions are
determined in a numerically exact manner, then the solution of the
eigenvalue problem is more than 2 orders of magnitude more time consuming
than the solution of the system of linear equations. In other words,
it becomes possible to compute more than 100 different geometries
in which the electromagnetic field is described by the same eigenvectors
and eigenvalues, with the same computing time needed for solving the
eigenvalue problem. This observation explains why it is possible to
perform a fairly exhaustive analysis of possible geometric structures
of lamellar type.

Clearly, when concrete experimental structures are to be modeled,
finite difference or finite element methods are more adequate than
such modal method that require some sort of stair-case approximation
to the measured geometry. However, if we wish to examine a wide variety
of structures for which the electromagnetic field can be expressed
in the same basis of eigenfunctions of the Helmholtz equation, then
the modal method based on eigenfunctions in the polynomial basis or
analytical eigenfunctions allows very fast calculation of whole spectra
for a large geometric parameter space. 

As typical structures used for PV solar cells comprise different absorbing
materials, e.g. Si, amorphous silicon (a-Si), doped ZnO, In doped
SnO$_{2}$ (often referred to as ITO), metals like Ag or Al, it is
important to be able to compute the absorption in a material specific
manner. For this purpose, we compute the directional energy flux density,
measured by the real part of the Poynting vector, along the borders
surrounding a particular material.

It turns out, that for metallic gratings, absorption losses in metallic
parts of the structure cannot in general be neglected, but for gratings
in which interfaces between metals and the semiconductor occur, these
losses can become dominant, such that the light absorption in the
semiconductor is severely affected. While in many cases the absorption
for the case in which the electric field points in the direction of
the grating grooves, referred to as E-polarization below, the absorption
in the metal is still quite small, the other case in which the magnetic
field points along the grating grooves, referred to as H-polarization
subsequently, leads to much larger or even dramatic parasitic absorption
in the metal, making such gratings useless for PV application.

In this this paper, we address the problem of parasitic absorption.
We show why the calculations in H-polarization are so much more difficult
than in E-polarization and what leads to slow convergence. We also
show for which types of gratings the parasitic absorption is reasonably
well behaved and also which types of structures must be avoided in
PV applications.

In a recent paper \citep{Gablinger:2014uq}, we have attempted to
give some limits for the absorption of light in solar cells assuming
an ideal antireflection coating and no absorption in the metal. Our
present paper reveals that this goal was ill-posed, because significant
absorption will always occur in metallic gratings, in particular for
light polarization for which the magnetic field has a component parallel
to metallic interfaces. For this reason, we have decided against publication
in a peer reviewed journal of that previous paper.

\section{Method}

In this section we first give a brief recapitulation of the calculational
method \citep{Morf:1995fi} that was used for the present calculations,
as well as a brief explanation on how the material specific absorption
is calculated.

\subsection{Setup}

\begin{figure}
\includegraphics[width=8cm]{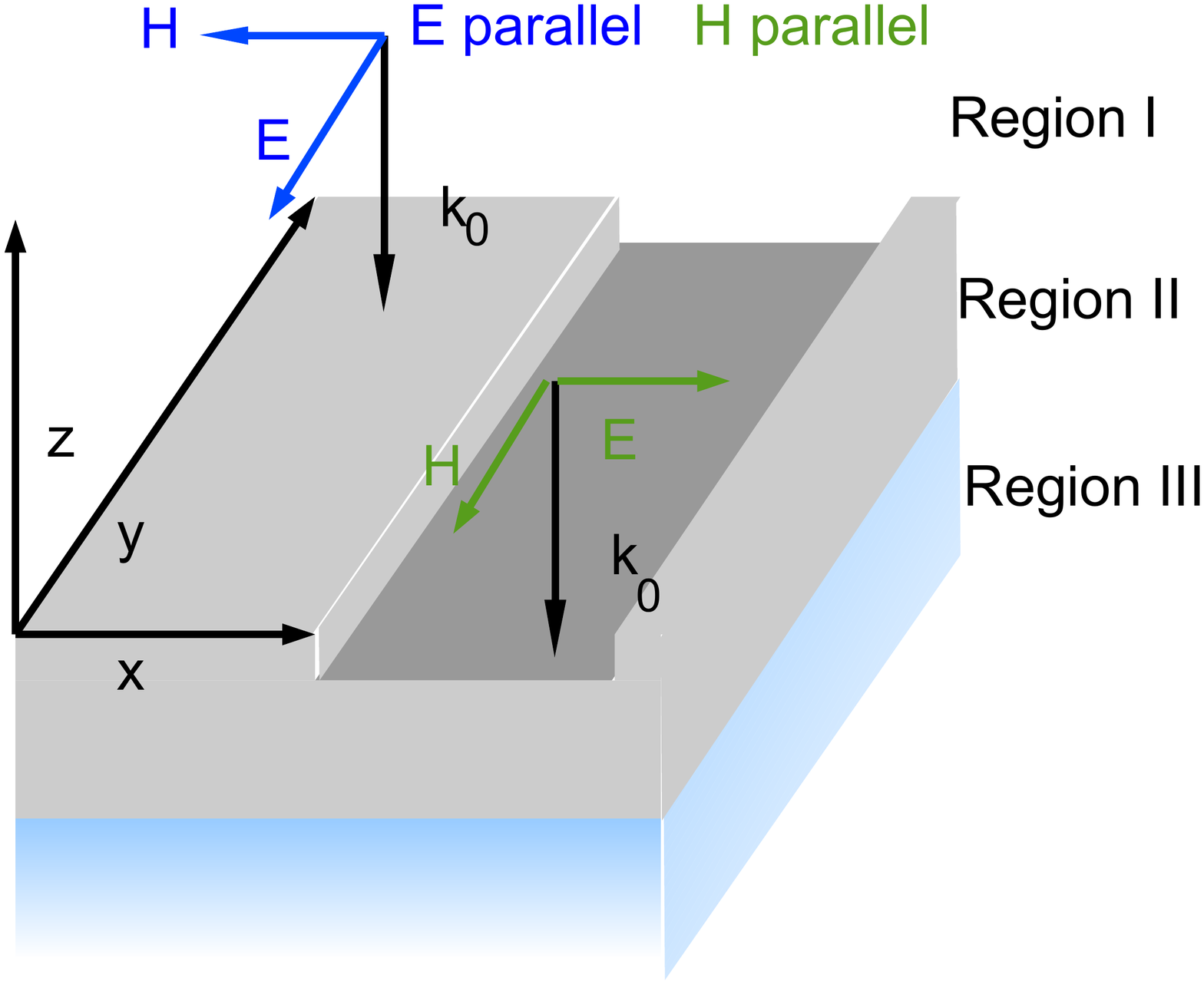}

\protect\caption{\label{coordinates}Problem setup and polarization convention.}
\end{figure}

We split the diffraction problem into regions I-III, as shown in Figure
\ref{coordinates}, using the polarization convention denoted there.
We denote $F(x,y)$ as either $E_{y}(x,y)$ for E parallel polarization
or $ $ In region I, we study the case of perpendicular incidence,
that is the incident light can be written as $F(x,y)=F^{inc}\exp[ik_{0}z]$,
and for both region I and III, the outgoing light can be written as
a superposition of plane waves, i.e. $F^{I}=\sum_{n=-M}^{M}a_{n}^{+}\exp[i(\alpha_{n}x+\chi z)]$
and $F^{III}=\sum_{n=-M}^{M}a_{n}^{-}\exp[i(\alpha_{n}x-\chi z)]$
respectively, where $\alpha_{n}=\frac{2\pi}{\Lambda}n$, and $\chi_{n}=\sqrt{\epsilon^{I,III}k_{0}^{\,2}-\alpha_{n}^{\;2}}$
. Region II contains the diffraction grating consisting of $q$ layers
, each $h_{j}=z_{j+1}-z_{j}$ thick, and where $j$ refers to the
layers position within the grating.

\subsection{Grating Problem}

As we confine our study to perpendicular incidence, we note that the
fields $E$ and $H$ do not depend on y. Within Region II, we briefly
note that within each layer the fields $F(x,z)$ can be written as
a sum of products $X(x))\times Z(z)$, which permits a separation
of variables, leading to the Helmholtz equation to be solved within
a grating layer. Because $\epsilon(x)$ is piecewise constant, the
Helmholtz equation can be written piecewise, where we limit ourselves
to two domains of constant permittivity. Hence,\ref{eq:helmholtz1}
refers to domain one and \ref{eq:helmholtz2} refers to domain two:

\begin{eqnarray}
\partial_{x}^{2}\, X_{n}^{(j,1)}(x)+k_{0}^{\;2}\epsilon_{1}^{(j)}X_{n}^{(j,1)}(x) & = & \lambda_{n}^{(j)\;2}X_{n}^{(j,1)}(x)\label{eq:helmholtz1}\\
\partial_{x}^{\;2}X_{n}^{(j,2)}(x)+k_{0}^{\;2}\epsilon_{2}^{(j)}(x)X_{n}^{(j,2)}(x) & = & \lambda_{n}^{(j)\;2}X_{n}^{(j,2)}(x)\label{eq:helmholtz2}
\end{eqnarray}

where $X_{n}^{(j,1)}$ and $X_{n}^{(j,2)}$ are the domain-wise eigenfunctions
and $\lambda_{n}^{(j)}$ are the eigenvalues of the Helmholtz equation.
The domain-wise eigenfunctions need to fulfill a set of boundary conditions
at interfaces $x_{i}$ in addition to the eigenvalue equations, that
is

\[
X_{n}^{(j,1)}(x_{i}^{-})=X_{n}^{(j,2)}(x_{i}^{+})
\]

\[
\partial_{x}X_{n}^{(j,1)}(x_{i}^{-})=\partial_{x}X_{n}^{(j,2)}(x_{i}^{+})
\]
 for E parallel polarized light, and 
\[
X_{n}^{(j,1)}(x_{i}^{-})=X_{n}^{(j,2)}(x_{i}^{+})
\]

\[
\frac{1}{\epsilon_{1}}\partial_{x}X_{n}^{(j,1)}(x_{i}^{-})=\frac{1}{\epsilon_{2}}\partial_{x}X_{n}^{(j,2)}(x_{i}^{+})
\]
 for $H$ parallel polarized light. 

With an eigenfunction $X_{n}^{(j)}$ composed of $X_{n}^{(j,1)}(x)$
and $X_{n}^{(j,2)}(x)$ we write the field as a sum of modes propagating
between the adjacent layers,

\begin{equation}
F^{(j)}(x,z')=\sum_{n=1}^{N}\left(a_{n}^{(j)-}\exp[+i\lambda_{n}^{(j)}z']+a_{n}^{(j)+}\exp[-i\lambda_{n}^{(j)}z']\right)X_{n}^{(j)}(x),\label{eq:planewave}
\end{equation}

where $z'=z-z^{(j)}$ .

Then the Helmholtz equation together with the sets of basisfunctions
and the additional constraint of boundary conditions between the patches
define the numeric eigenvalue problem for the layer $j$ to be solved.
This method allows to calculate a freely selectable number of eigenvalues
and eigenfunctions to a precision limited by machine number size because
of its exponential convergence. That is, within a grating layer, our
eigenmodes are obtained to the same precision as in an analytic calculation
following references by \citet{Botten:1981jb,Botten:1981kq,Botten:1981tp}.
Note that the symmetry at perpendicular incidence together with the
symmetry within the grating layer result in either purely symmetric
or antisymmetric functions, that is we only need half the number of
variables, and therefore the symmetric eigenvalue problem requires
only $\frac{1}{8}$ of the numerical effort as it scales with the
number of variables to the third power, and for symmetric structures,
this property carries over to the solution of the radiation condition
boundary value problem. To this end, we follow \citep{Morf:1995fi}
and, in addition, by using a minimum bandwidth banded matrix implementation
that will be explained in detail in an upcoming publication.

At the interface $z=z_{0}$ between grating layers, the field $F=E_{y}$
or $F=H_{y}$ has to satisfy the following continuity equations
\[
F(x,z_{0}+\delta)\equiv F(x,z_{0}-\delta)\text{ }
\]

\[
\partial_{z}F(x,z_{0}+\delta)\equiv\partial_{z}F(x,z_{0}-\delta),
\]

and the fact that within each layer there exists a different set of
eigenmodes, means that a high number of modes have to be used to calculate
the behaviour at such interfaces - regardless of the numeric method
used. 

The substantial reduction in modes through symmetry together with
the implementation of the boundary value problem allows us to achieve
the high resolution we need to describe both the spectra of such structures
and their absorption properties.

\subsection{Material Specific Absorption\label{sub:Local-Absorption}}

\begin{figure}[h]
\includegraphics[clip,width=0.65\columnwidth]{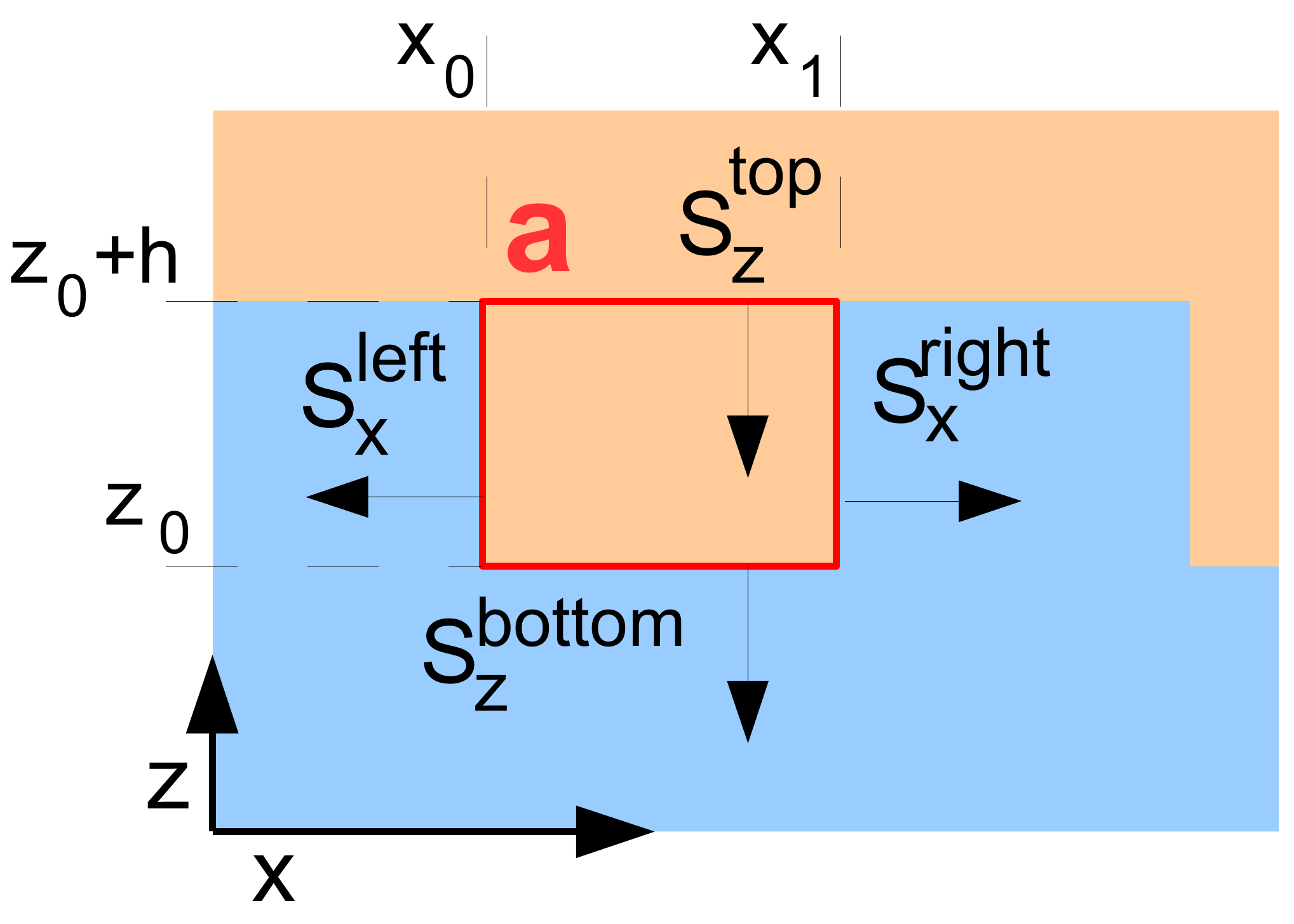}

\protect\caption{\label{fig:PoyntingSchematic}schematic for the absorption within
an area A }
\end{figure}

To calculate the absorption within a lamellar slab, we make use of
the Poynting vector along the boundary of a region containing the
material for which the absorption needs to be known. The time-average
of the Poynting Vector is given by \citep{Born:1999zp}

\[
\vec{S}=\frac{c}{8\pi}\Re\vec{[E}\times\vec{H}^{*}].
\]
It represents the directional energy flux density of the electromagnetic
field. 

Assuming the Poynting vector $\vec{S}\left(\vec{r}\right)$ is known
for all $\vec{r},$ the absorption $A$ in an area $a$ with border
$\partial a$ becomes $A=\int_{\partial a}\vec{S}\left(\vec{r}\right)\textrm{d}r$.
For a one-dimensional grating and non-conical incidence in the xz-plane,
the electromagnetic field does not depend on $y$ and the integration
in the y-direction can be omittted. For both polarization, the Poynting
vector will have non-zero $x-$ and $z-$components, i.e. $\vec{S}=\left(S_{x},0,S_{z}\right)$.
The absorption $A$ in the area $a$ can then be written as 
\[
A=\int_{z_{o}}^{z_{0}+h}\left(S_{x}^{left}\left(z\right)+S_{x}^{right}\left(z\right)\right)\textrm{d}z+\int_{x_{0}}^{x_{1}}\left(S_{z}^{top}\left(x\right)+S_{z}^{bottom}\left(x\right)\right)\textrm{d}x
\]

For E (H)-polarisation, only$E_{y}$ ($H_{y})$ is non-zero. Using
Maxwell's equation, the Poynting vector can then be written as 
\begin{eqnarray*}
\vec{S} & = & \vec{e}_{x}\frac{i}{k_{0}\mu}E_{y}\frac{\partial E_{y}^{*}}{\partial x}-\vec{e}_{z}\frac{i}{k_{0}\mu}E_{y}\frac{\partial E_{y}^{*}}{\partial z}\\
\vec{S} & = & \vec{e}_{z}\frac{i}{k_{0}\epsilon}\frac{\partial H_{y}}{\partial z}H_{y}^{*}-\vec{e}_{x}\frac{i}{k_{0}\epsilon}\frac{\partial H_{y}}{\partial x}H_{y}^{*}
\end{eqnarray*}

for E- and H-polarization, respectively. For both polarization the
field $F(x,z)=E_{y}(x,z)$ or $F(x,z)=H_{y}(x,z)$ and its derivative
can be written as

\begin{eqnarray*}
F & = & \sum\left(a_{n}^{+}e^{i\mu_{n}z}+a_{n}^{-}e^{i\mu_{n}\left(h-z\right)}\right)X_{n}\left(x\right)\\
\frac{\partial F}{\partial x} & = & \sum\left(a_{n}^{+}e^{i\mu_{n}z}+a_{n}^{-}e^{i\mu_{n}\left(h-z\right)}\right)\frac{\partial X_{n}\left(x\right)}{\partial x}\\
\frac{\partial F}{\partial z} & = & \sum i\mu_{n}\left(a_{n}^{+}e^{i\mu_{n}z}-a_{n}^{-}e^{i\mu_{n}\left(h-z\right)}\right)X_{n}\left(x\right)
\end{eqnarray*}

Without loss of generality, the coordinate $z_{0}$ can be set to
$z_{0}=0$. Then, at $x_{1}$ or $x_{2}$, the integrated contribution
of the Poynting vector in $z$-direction becomes

\begin{eqnarray*}
\frac{-i}{k_{0}\epsilon}\int_{0}^{h}\frac{\partial H_{y}}{\partial x}H_{y}^{*}\textrm{d}z & = & \frac{-i}{k_{0}\epsilon}\int_{0}^{h}\sum_{n,m}C_{n,m}\left(a_{n}^{+}e^{i\mu_{n}z}+a_{n}^{-}e^{i\mu_{n}\left(h-z\right)}\right)\times\\
 & \times & \left(a_{m}^{+}e^{i\mu_{m}z}+a_{m}^{-}e^{i\mu_{m}\left(h-z\right)}\right)^{*}dz\\
 & = & \frac{-1}{k_{0}\epsilon}\sum_{m,n}C_{n,m}\Biggl(\left(a_{n}^{+}a_{m}^{+*}+a_{n}^{-}a_{m}^{-*}\right)\frac{\left(e^{ih\left(\mu_{n}-\mu_{m}^{*}\right)}-1\right)}{\mu_{n}-\mu_{_{m}}^{*}}+\\
 & + & \left(a_{n}^{+}a_{m}^{-*}+a_{n}^{-}a_{m}^{+*}\right)\frac{\left(e^{i\mu_{n}h}-e^{-i\mu_{m}^{*}h}\right)}{\mu_{n}+\mu_{m}^{*}}\Biggr).
\end{eqnarray*}

where

\[
C_{n,m}=\Bigl(\frac{\partial X_{n}}{\partial x}X_{m}^{*}\Bigr)(x_{k})
\]

is evaluated at the boundaries of the corresponding domain $x_{1},\, x_{2}$,
respectively. Here and in the following, we drop the common factor
$c/8\pi$ in the expression for the Poynting vector. Since all absorption
results will be quoted as a ratio with respect to the Poynting vector
of the incident radiation, this factor would drop out anyway.

The contributions from the Poynting vector in z-direction at the top
$S_{z}^{top}$ and bottom $S_{z}^{bottom}$ integrated with respect
to $x$, are quite easy to compute since the overlaps $C'_{n,m}=\int_{x_{\text{1}}}^{x_{2}}\left(X_{n}X_{m}^{*}\right)\textrm{d}x$
can be expanded in terms of products of Legendre polynomials that
are orthogonal when integrated over a given domain consisting of the
material for which the absorption has to be computed.

Thus only diagonal terms in the Legendre expansion of $X_{n}$ and
$X_{m}$ survive, and the corresponding matrix $C_{n,m}$ can be evaluated
as soon as the Legendre expansion coefficients of the eigenfunctions
are known. Note how much more cumbersome such a calculation would
be if instead of the Legendre basis a Fourier basis had been used.
Then no orthogonality relation would exist for a single domain.

\section{The set of geometric structures}

Here, we wish to present the structures which will be calculated in
the following sections. The goal of these calculations will be to
obtain the spectra of absorption in amorphous silicon and in the other
absorbing materials, i.e. silver, aluminum, zincoxide and indium tin
oxide, where absorption does not contribute to the generation of electron-hole
pairs and should be kept as small as possible.

The first two structures are following the standard structure for
amorphous silicon solar cells. They are formed by deposition of zinc
oxide on a metallic reflector. Subsequently, amorphous silicon is
deposited and last an indium tin oxide layer is deposited as a transparent
contact. We ignore the fact that the amorphous silicon layer consists
of three parts, n-doped - intrinsic and p-doped and consequently the
absorption in the doped parts, that will not contribute to the photovoltaic
efficiency.

\begin{figure}[h]
a)\includegraphics[width=5cm,height=4.7cm]{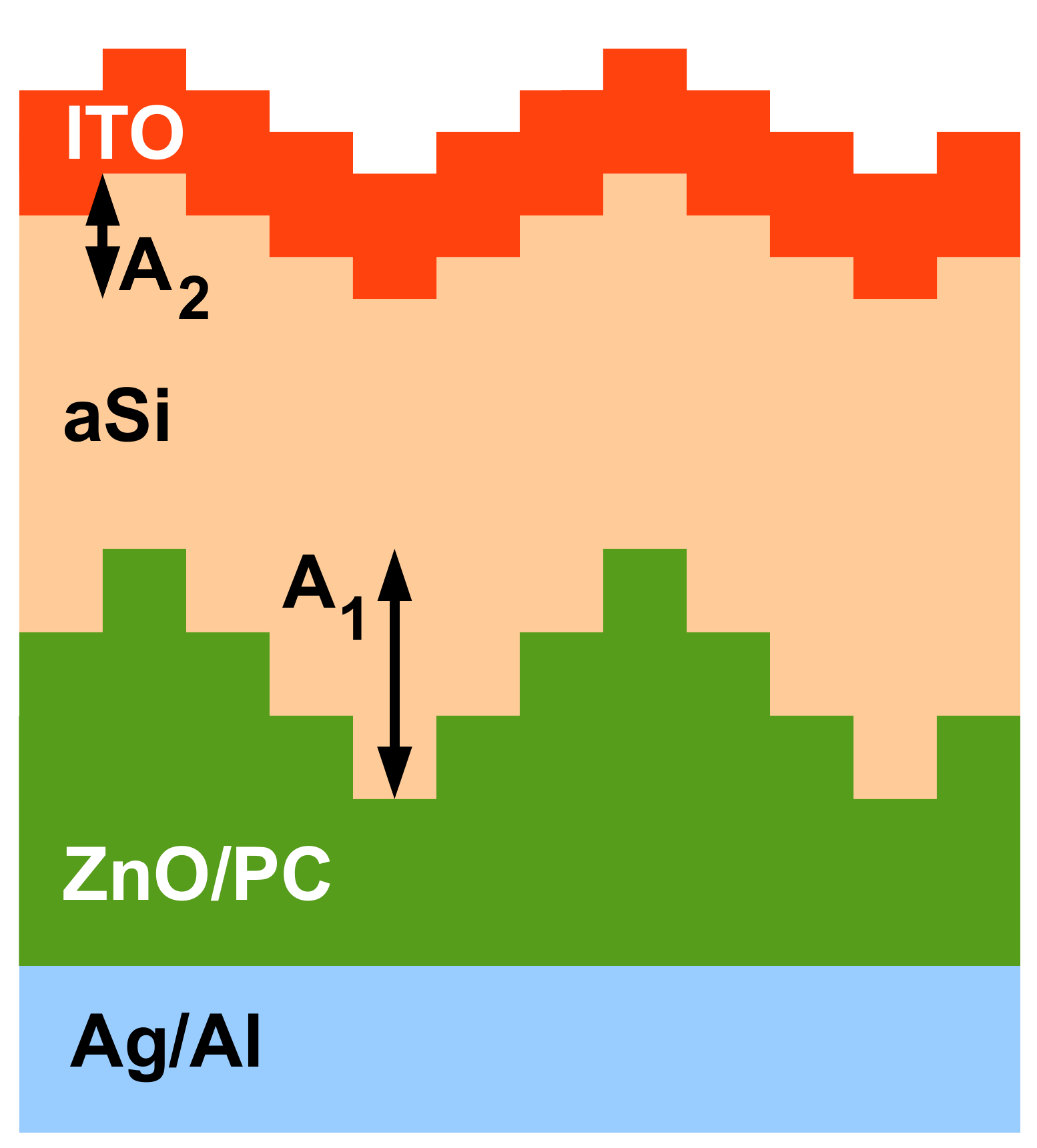}b)\includegraphics[width=5cm]{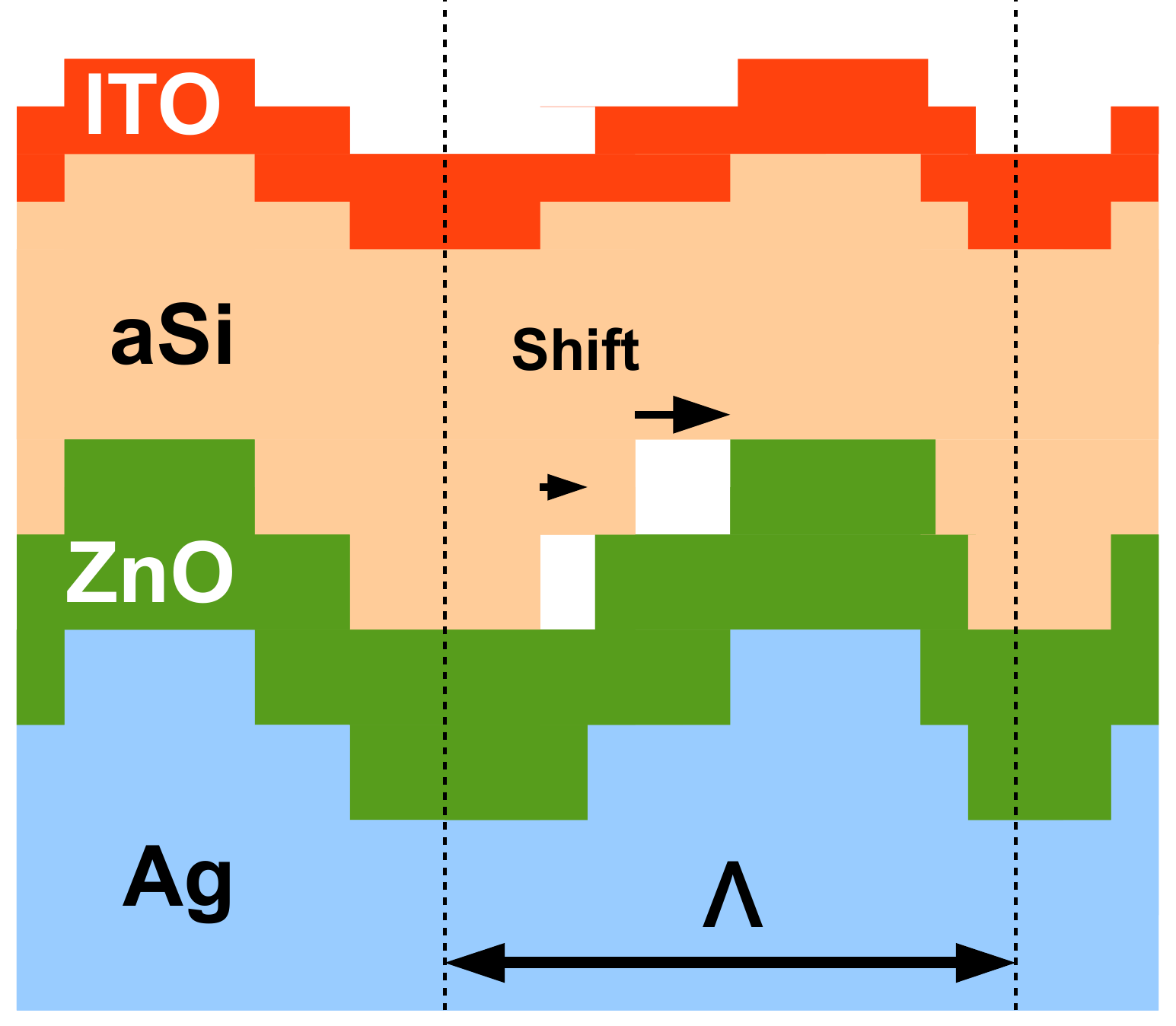}

c)\includegraphics[width=5cm,height=4.7cm,keepaspectratio]{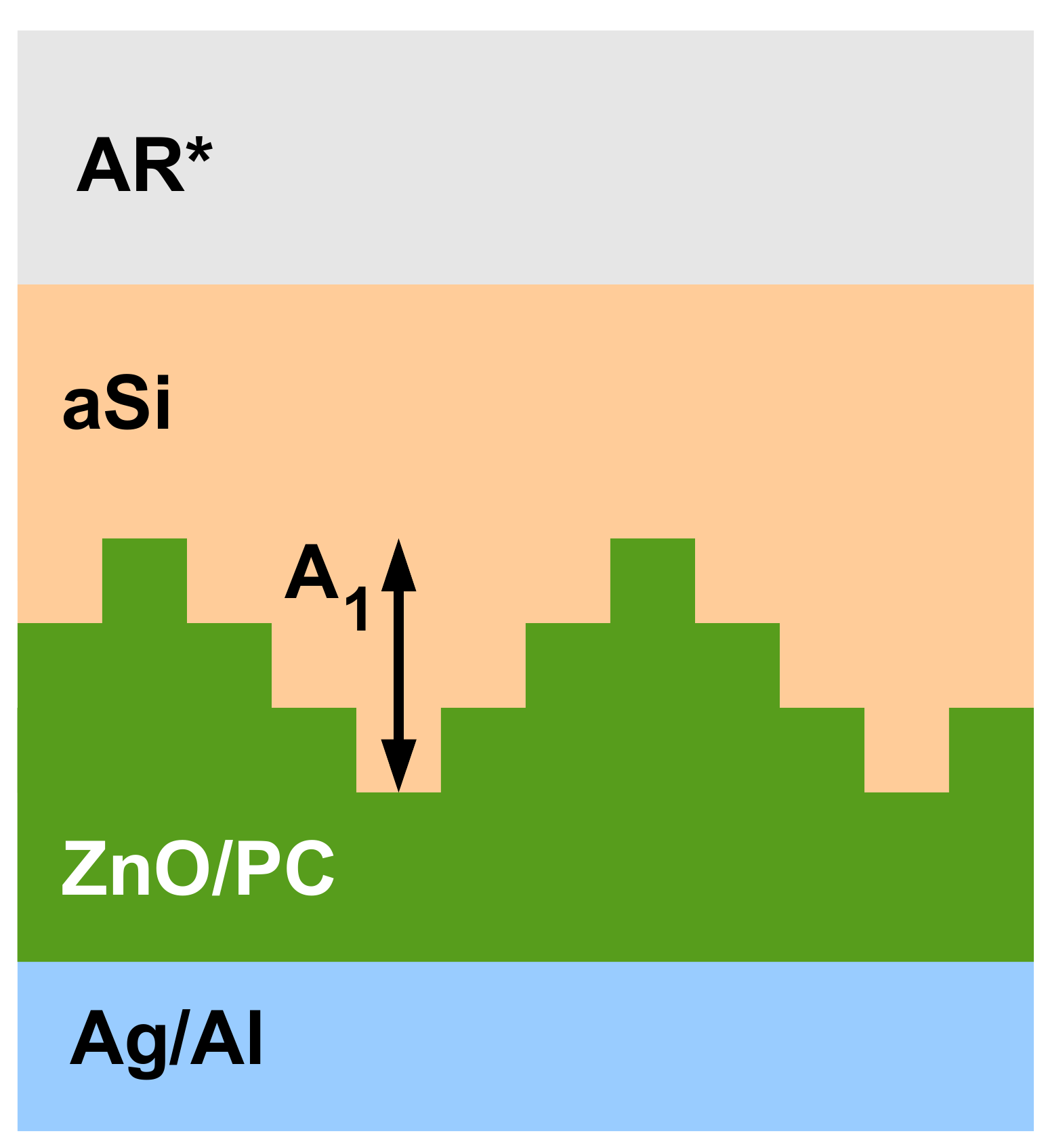}d)\includegraphics[width=5cm,height=4.7cm,keepaspectratio]{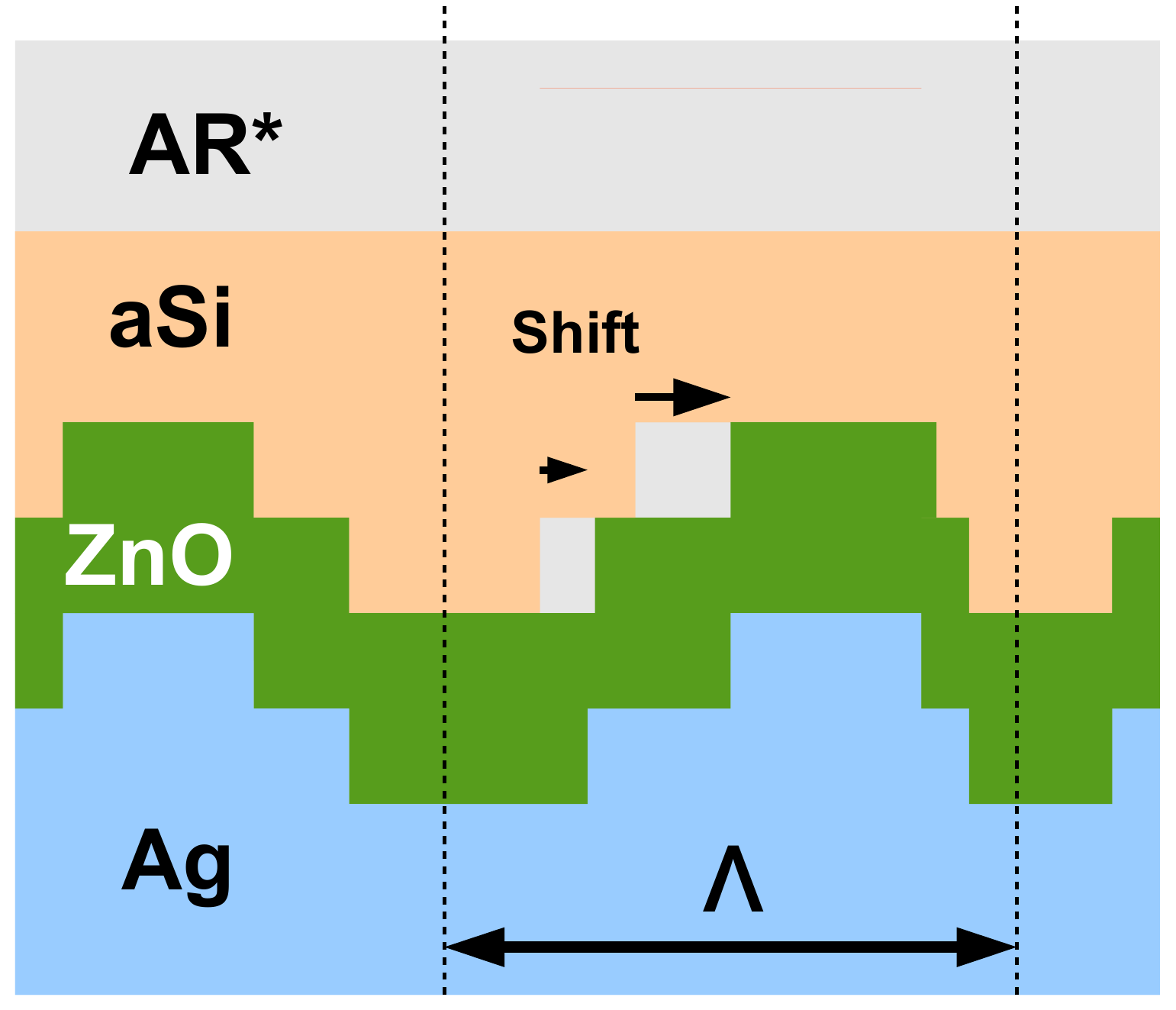}

\protect\caption{The 4 types of structures studied in this work. In the terminology
introduced below, the structures in panels a) and c) belong to the
$P^{\left(3\right)}$ class, while panels b) and d) belong to $P^{\left(2\right)}$.\label{fig:The-4-types of structures}}
\end{figure}

In the first two structures, Fig. 3a and 3b, the grating structure
of the substrate, Ag in Fig. 3a and ZnO or polycarbonate in Fig. 3b,
is reproduced in all subsequent layers which are deposited on the
substrate, albeit with a somewhat reduced amplitude above the homogeneous
a-Si layer, leading to some antireflection effect at the top of the
structure. Fig. 3c and 3d show structures with a planar top a-Si surface
and some artificial antireflective coating AR{*} which suppresses
reflection. It is used in order to confine the investigation to the
optimisation of light-trapping, cf Appendix, and study the benefits
and deficiencies of structures 3c and 3d.

In addition, we also show how we approximate smooth surface relief
gratings for the case of a sinusoidal profile in Fig. 4. For illustration
purpose, we show 3 and 5 step versions. In this construction, we make
use of a variable step height and step position in the horizontal
direction, such that excess areas (yellow triangles above the sine)
and missing areas (blue triangles below the sine) compensate exactly
and at the same time the sum of the excess areas is minimized, while
the sum of missing areas is minimized as well. This leads to a unique
solution for any given number of steps, maintains the symmetries and
the amount of material exactly. We will see below, that for increasing
number of steps, the absorption spectra change very little.

For brevity, we will denote structures consisting of combinations
of such materials and gratings as $\mbox{Mat}_{1}|P^{\left(n\right)}\left(\mbox{Mat}_{2}|\mbox{Mat}_{3}\right)|\mbox{Mat}_{4}$,
where we start from left with the substrate, and where the pseudosine
$P^{\left(n\right)}$ contains the materials $\mbox{Mat}_{2}$ and
$\mbox{Mat}_{3}$.

\begin{figure}[H]
\includegraphics[clip,width=6cm,height=2.5cm,keepaspectratio]{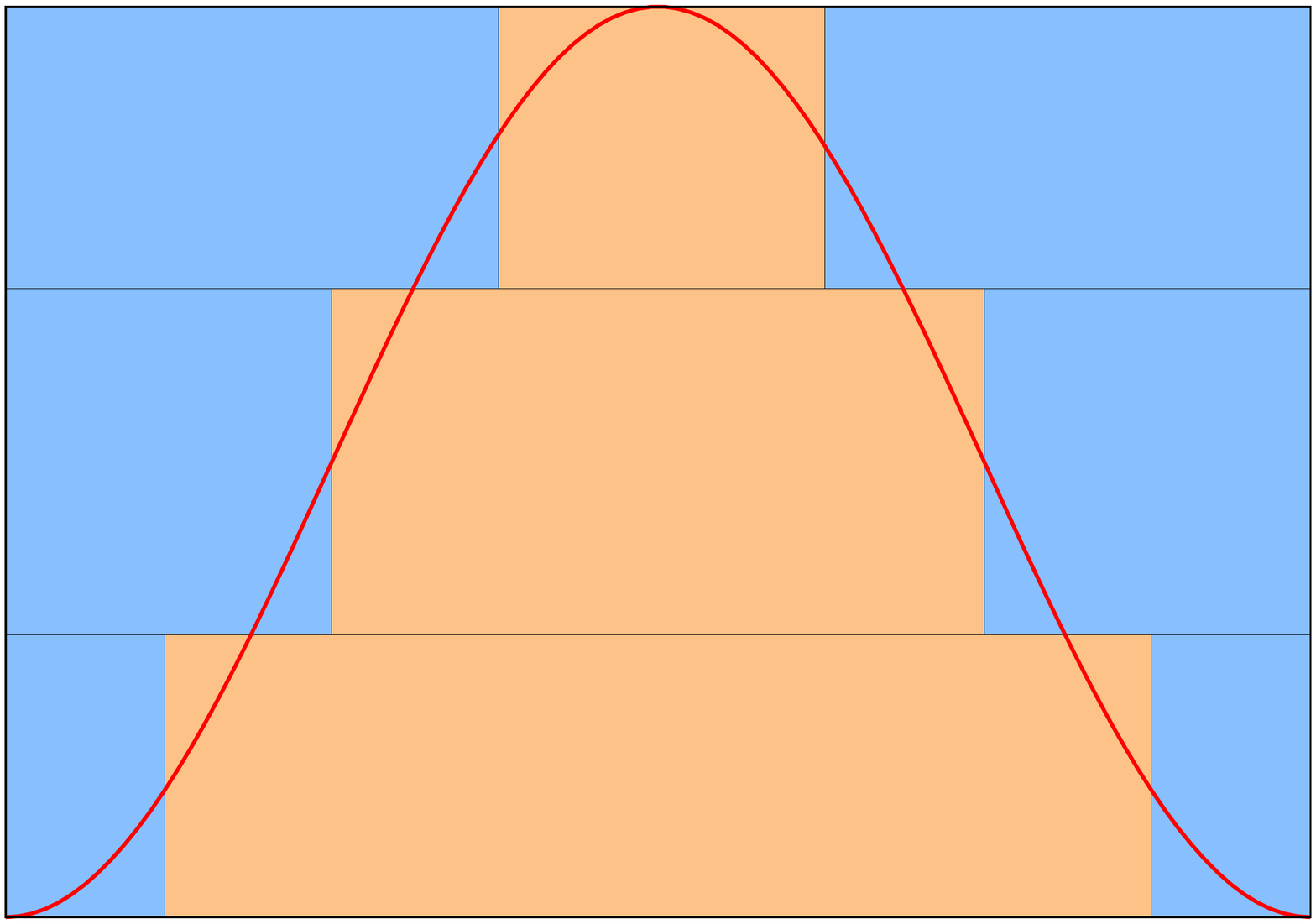}\includegraphics[bb=0bp 0bp 670bp 472bp,clip,width=6cm,height=2.5cm,keepaspectratio]{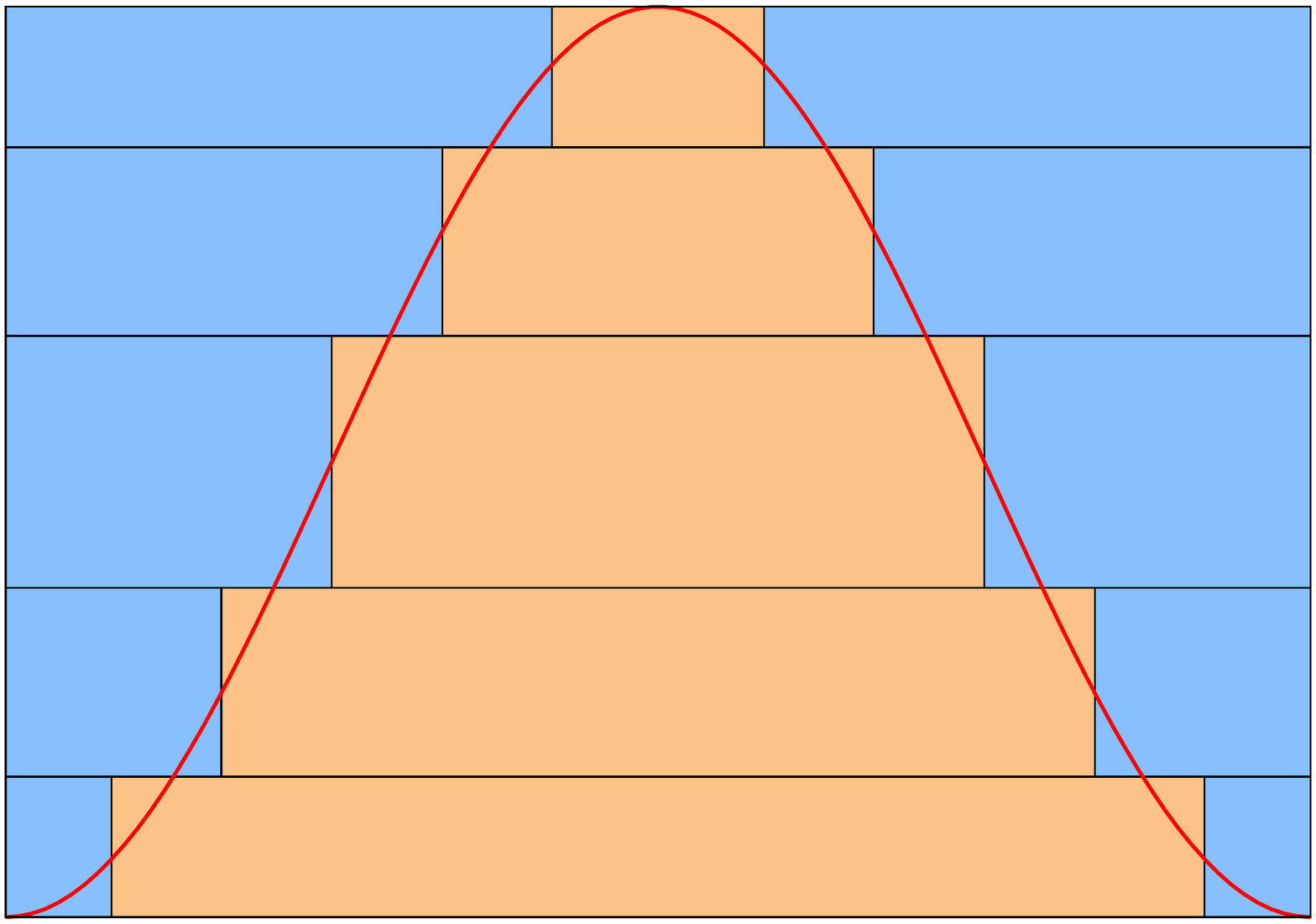}

\protect\caption{Mimicking a sine-grating by a stair-case structure with 3 and 5 steps.
We will refer to such structures as pseudo sine of types $P^{\left(3\right)}$
and $P^{\left(5\right)}$.\label{fig:Mimicking-a-sine-grating}}
\end{figure}

\section{Results}

\paragraph*{Spectrally Integrated Absorption}

For our calculations of integrated absorption, the following reference
data should be kept in mind: The photon AM1.5 spectrum integrated
from 380-770 nm counts as 100\%, and a flat structure with a silver
back reflector covered by an optimal layer of 50nm ZnO, 200nm a-Si
and 60nm ITO on top absorbs 51\% in the a-Si layer at perpendicular
incidence. The contribution integrated from the wavelength range 550-770nm
is about 19\%. The cut-off wavelength of 380nm is determined both
by the solar spectrum AM1.5 and by the very small decay length for
photons below 400nm wavelength, such that the quantum efficiency becomes
quite small. The upper cut-off wavelength of 770nm must be regarded
as an upper limit for amorphous silicon whose band gap is between
1.6 and 1.7 eV, unless it is alloyed with germanium as is used in
some amorphous silicon tandem cells.

\begin{table}

\begin{tabular}{|c|c|c|}
\hline 
spectral
range & a-Si|AR{*} & Ag|a-Si|AR{*}\tabularnewline
\hline 
380-770nm & 0.97727 & 0.55247\tabularnewline
\hline 
380-550nm & 0.35589 & 0.35241\tabularnewline
\hline 
550-770nm & 0.62138 & 0.20006\tabularnewline
\hline 
600-770nm & 0.48215 & 0.08891\tabularnewline
\hline 
\end{tabular}\protect\caption{Reference data for a structure in which the reflection at the front
surface is suppressed. If all photons in the specified spectral range
are absorbed and converted with quantum efficiency 1, the AM1.5-weighted
integrated absorption within a-Si with an artificial idealized antireflection
coating AR{*} is listed in the second column. The third column lists
the data for a 200nm a-Si layer with AR{*} coating on a silver substrate.
AR{*} leaves a small residual reflection of 0.00339 in the spectral
range from 380-770nm.}
\end{table}

\subsection{Validity of our calculations}

\subsubsection{Convergence of spectra for increasing number of modes}

In this section, we present a few tests concerning the reliability
and validity of our calculational method.

\begin{figure}[h]
\includegraphics[width=0.5\columnwidth]{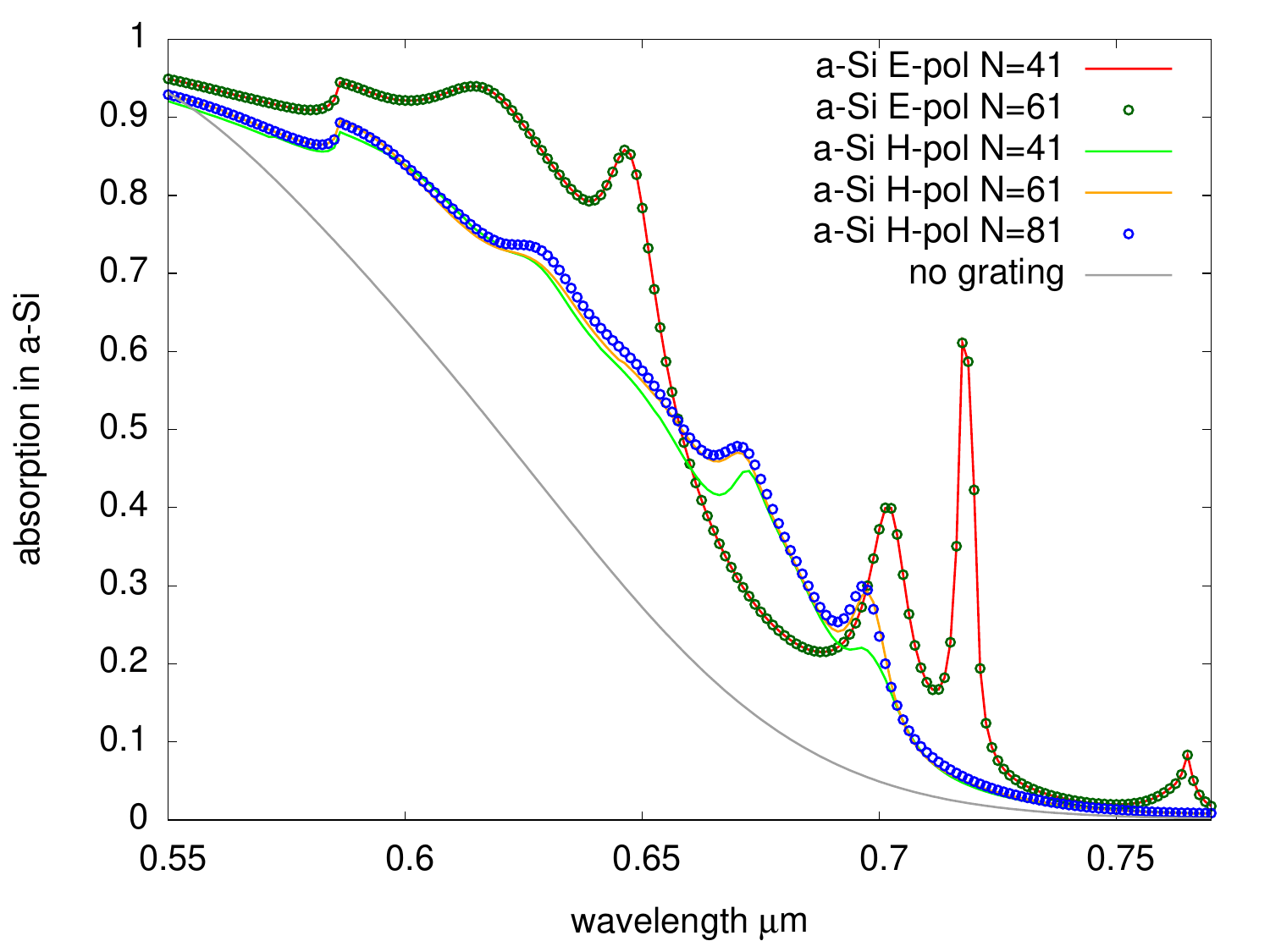}\protect\caption{The modal convergence of the spectrum of a grating of type $P^{\left(7\right)}\left(\mbox{Ag|a-Si}\right)|\mbox{AR}^{*}$\label{fig:The-modal-convergence}}
\end{figure}

The first test is with a 7-step grating mimicking a sine grating (``7-step
pseudo-sine'') \ref{fig:Mimicking-a-sine-grating} in a silver substrate
with a total amplitude (peak to valley) of 64nm and a grating period
of 586nm. Figure \ref{fig:The-modal-convergence} we show that even
the difficult case of a 7-step metallic pseudo-sine grating can be
calculated reliably using a limited number of modes. For E-polarization,
using 41 numerically exact modes are sufficient to calculate the spectrum
accurately. For H-polarization, 61 numerically exact modes are required
for H polarization, where the spectrum becomes virtually indistinguishable
from a spectrum with 81 modes. Note the small deviation in the result
for H-polarization around 630-670nm and close to 700nm wavelength.
We will see below, that the absorption in silver is less well behaved.

\subsubsection{Energy conservation\label{sub:Energy-conservation}}

Another interesting test is provided by the requirement of energy
conservation. Indeed, the sum of absorption in the different absorbing
materials together with total reflection and transmission probabilities
has to be close to unity if the calculations can be trusted. Here,
we show the results of a particularly difficult calculation consisting
of a square wave grating in a Ag substrate with height 60nm onto which
an amorphous silicon layer of 200nm thickness is conformally deposited,
thereby leading to a second square wave grating in Si and vacuum with
the same height \ref{fig:The-4-types of structures}a) without ZnO
and ITO.

\begin{figure}
\includegraphics[clip,width=6cm]{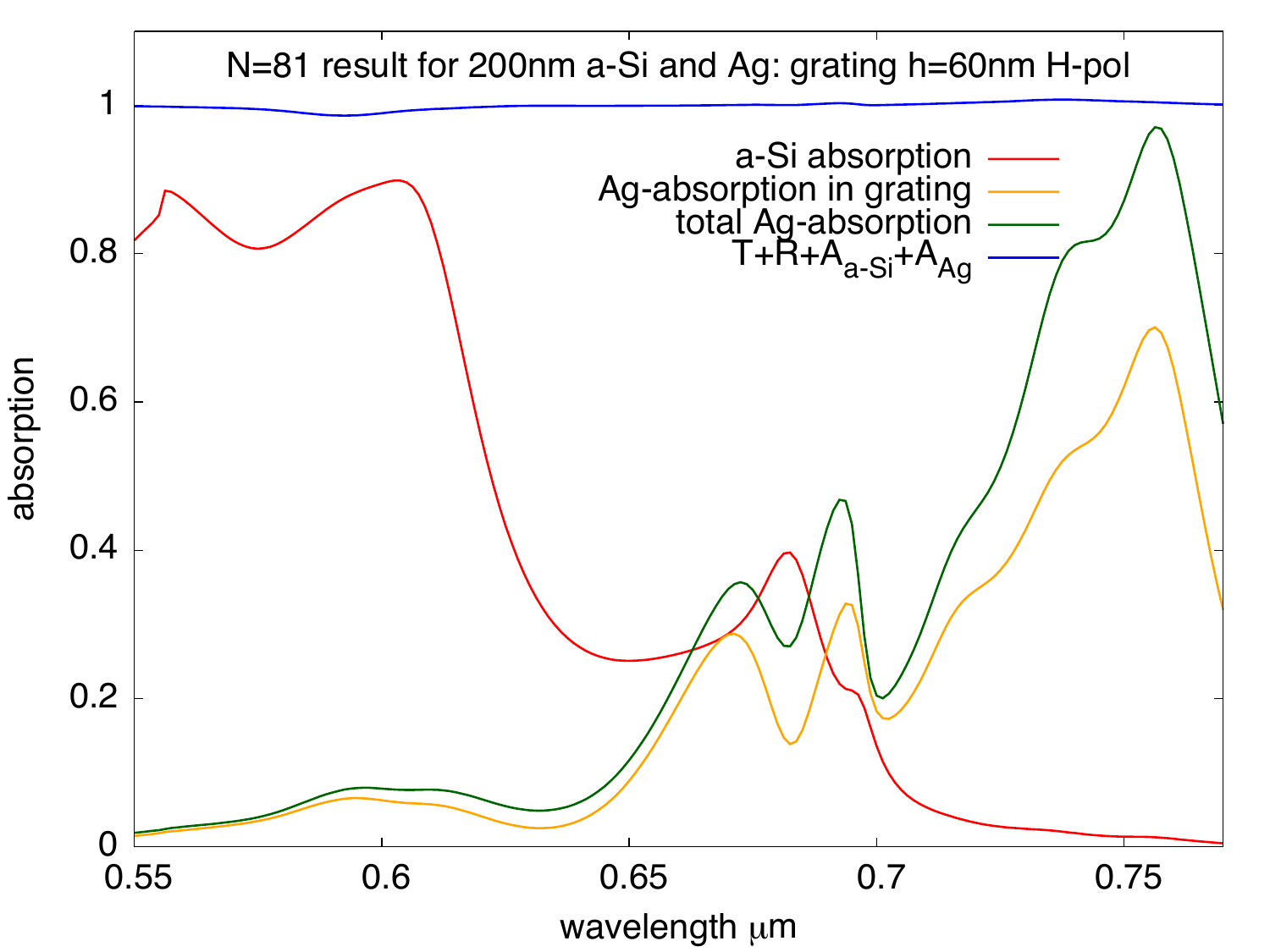}\includegraphics[clip,width=6cm]{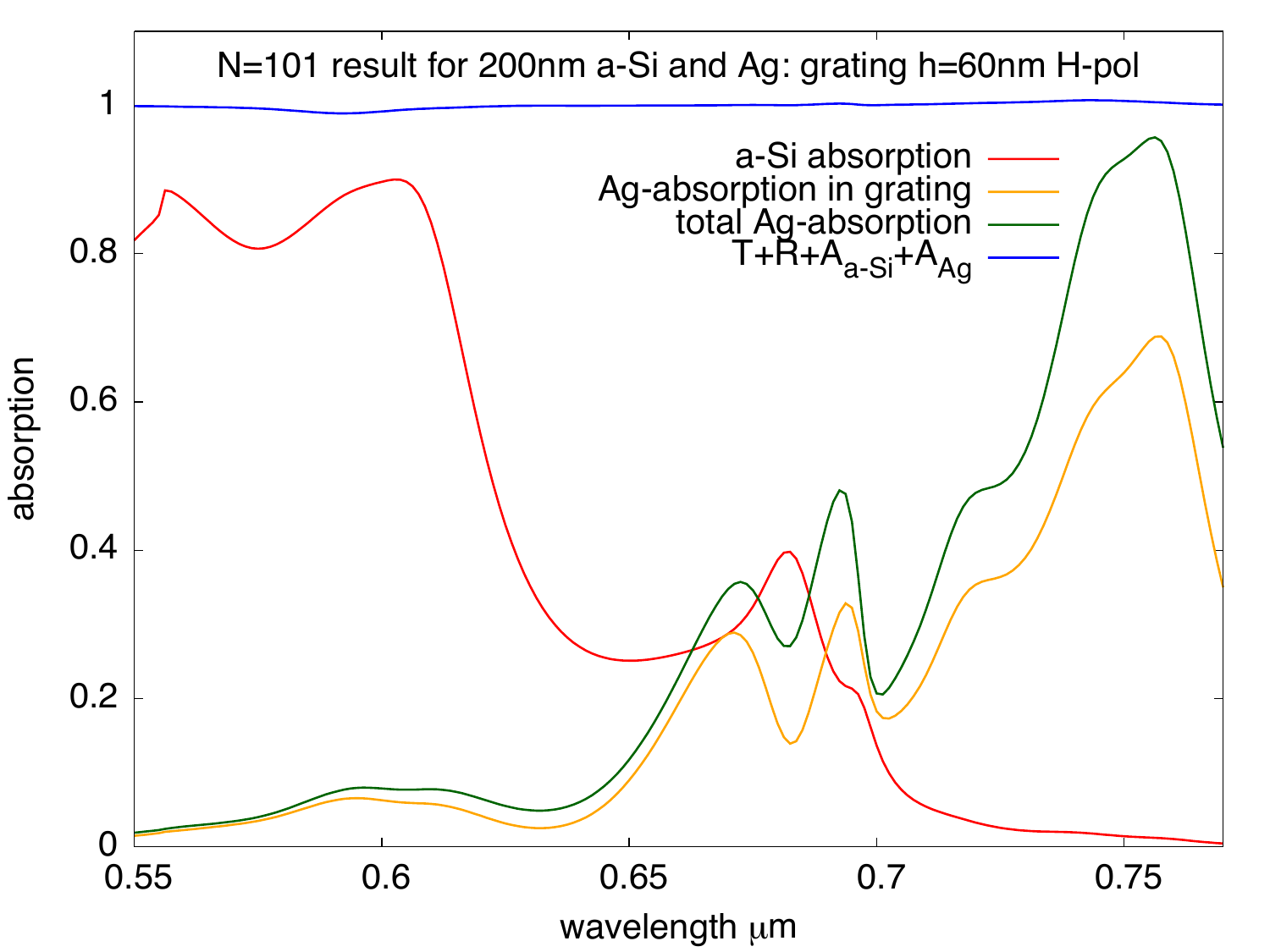}

\protect\caption{Absorption spectrum in H-polarization of a (rectangular) grating of
type $P^{\left(1\right)}\left(\mbox{Ag|a-Si}\right)$|a-Si| $P^{(1)}$(a-Si|
air) with period 556.1nm and grating depth 60nm, using N=81 (left
panel) and N=101 modes (right panel). The effective a-Si thickness
is 200nm. Shown are the spectrum of absorption in A(a-Si) in silicon
and A(Ag) in silver, separately, as well as the sum T+R+A(a-Si)+A(Ag)
as a test of energy conservation.\label{fig:Spectrum-in-H-polarisation square Ag-Si grating}}
\end{figure}

In Figure \ref{fig:Spectrum-in-H-polarisation square Ag-Si grating}
we observe a number of interesting features: (i) Huge parasitic absorption
in the Ag grating for wavelengths above 700nm (yellow line) and also
a growing contribution from the substrate absorption (green line).
By contrast, in this spectral range the absorption in a-Si is very
weak. (ii) Comparing the results of N=81 and N=101 modes, we observe
that the main difference occurs in the absorption of Ag above about
720nm wavelength. This observation is consistent with the findings
of Popov et al. \citep{Popov:2002jf}that the convergence of stair-case
approximations to smooth metal surfaces is poor or even a serious
problem. (iii) The spectrum of absorption within the amorphous Si
(red line) is basically unchanged when the number of modes is increased.
(iv) The energy conservation (blue line) is very stable and close
to unity even at wavelengths where resonances are visible.

For the same structure, we also show the results for E-polarization.
In this case, the differences between N=81 and N=101 modes are invisible
on the scale of the plot. Both absorption in a-Si and Ag are quite
a bit lower that in H-polarization. The parasitic absorption in Ag
shows much narrower resonances, presumably due to much coupling between
different modes. Energy conservation is even better behaved than in
H-polarization in spite of the strong and sharp absorption resonances.

\begin{figure}
\includegraphics[clip,width=6cm]{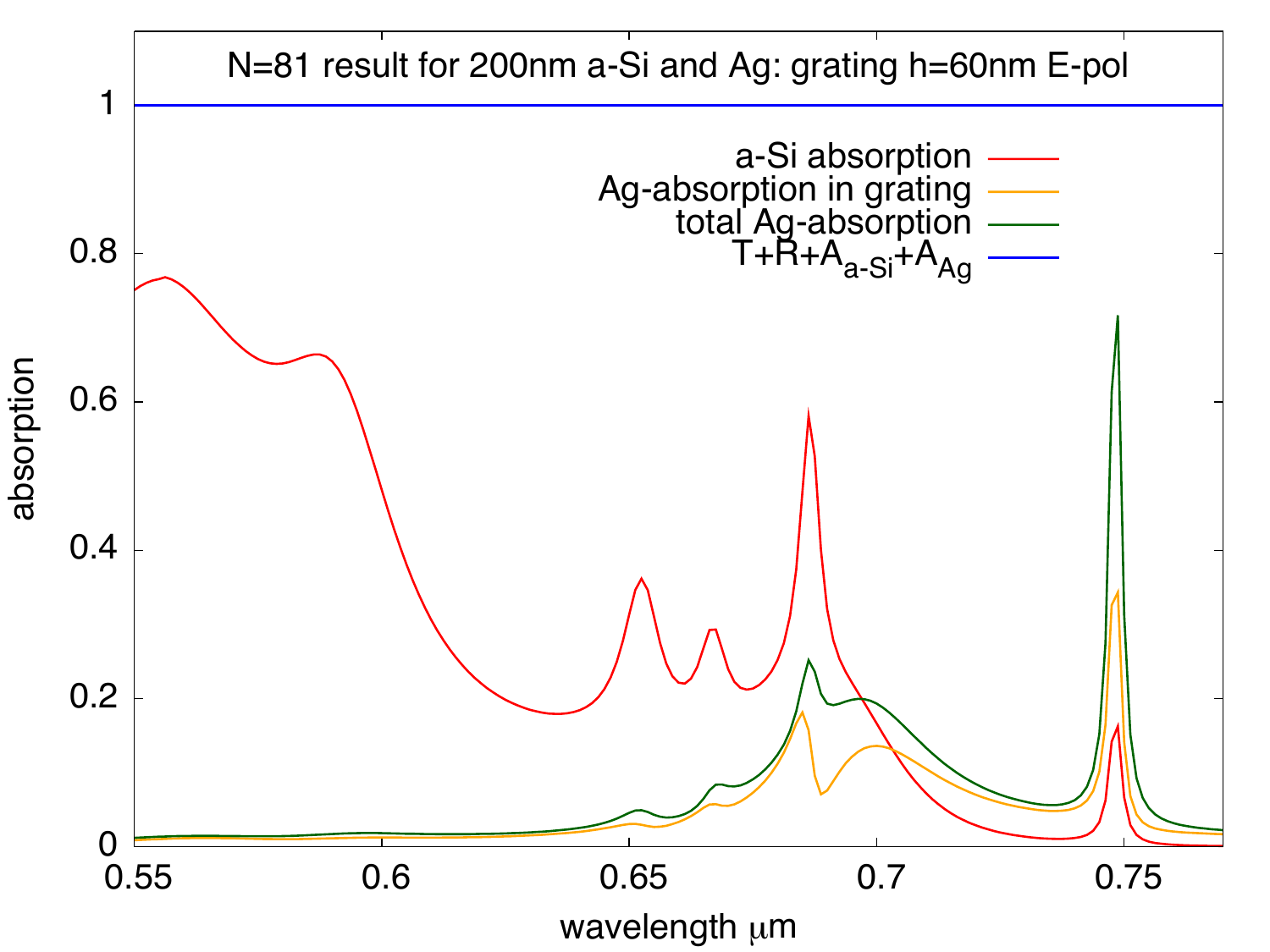}\protect\caption{Spectrum in E-polarization for the same structure as in Fig. \ref{fig:Spectrum-in-H-polarisation square Ag-Si grating}.}
\end{figure}

\subsubsection{Sinusoidal gratings mimicked by stair case gratings:\protect \\
convergence in the number of steps}

\begin{figure}
\includegraphics[clip,width=6cm]{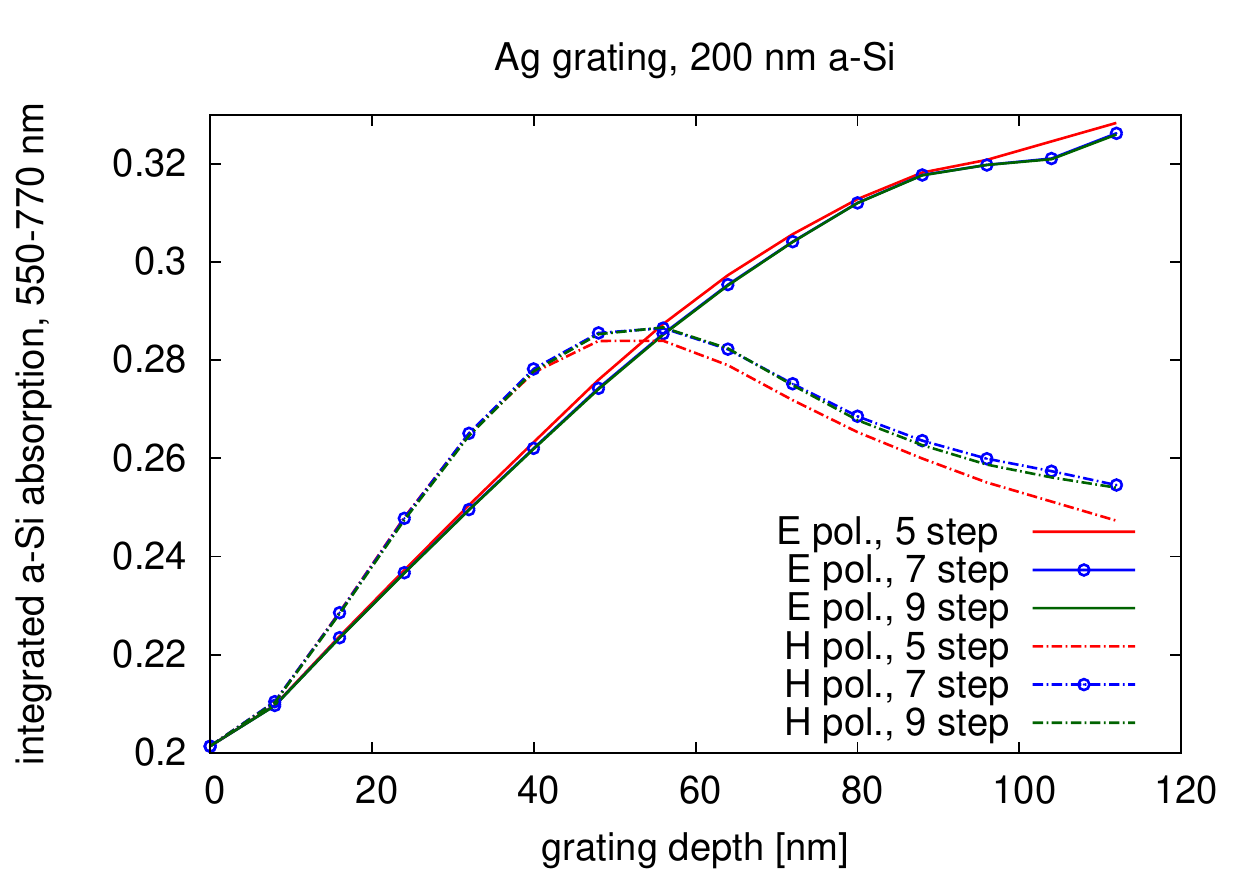}\includegraphics[clip,width=6cm]{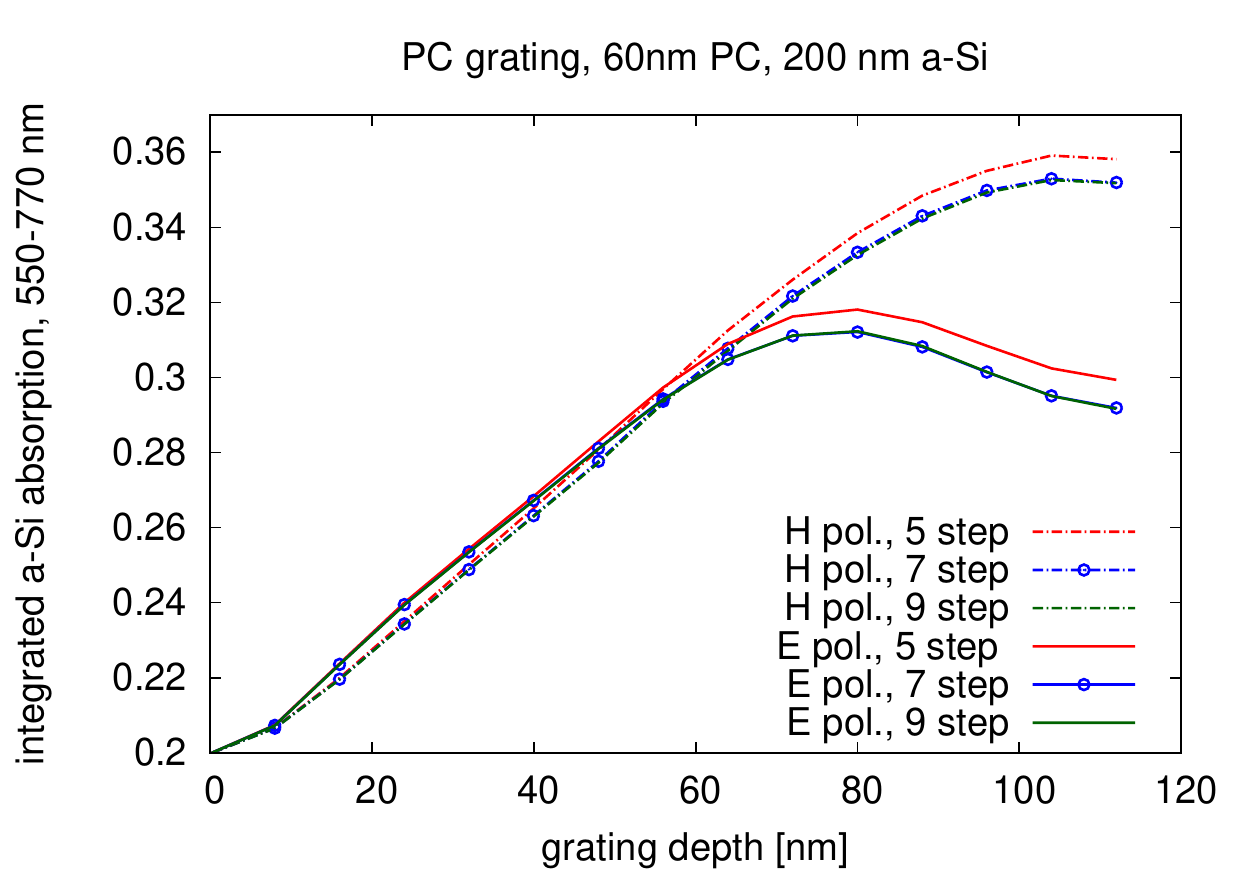}\protect\caption{Integrated over 550nm$<\lambda<770$nm, AM1.5-weighted, absorption
in a-Si: grating in the silver substrate (left panel), grating in
a PC overcoat of the silver substrate shown as a function of the total
height of the grating (right panel). Comparison of different types
of $P^{\left(n\right)}.$ Left panel: Structures of types Ag$|P^{\left(5\right)}-P^{\left(9\right)}\left(\mbox{Ag|a-Si}\right)|\mbox{AR}^{*}$Right
panel: Structures of types $\mbox{Ag}|\mbox{PC}|P^{\left(5\right)}-P^{\left(9\right)}\left(\mbox{PC|a-Si}\right)$AR{*}.\label{fig:Pseudosine-gratings-step-conv}}
\end{figure}

We now turn our attention to the behavior of the absorption spectra
of ``pseudosine'' gratings, when the number of steps is increased.
The structure corresponds to the one of Fig.\ref{fig:The-4-types of structures}d)
with a grating shape of \ref{fig:Mimicking-a-sine-grating}. The spectral
results \ref{fig:Pseudosine-gratings-step-conv} show that both for
the grating in silver (left panel) as well as for the grating in PC,
the results are quite insensitive to the number of steps even for
H-polarization and a grating with the dangerous interface between
silver and amorphous silicon. Indeed, the difference between 7 and
9 steps is negligeable in all cases, while the 5-step result shows
the same qualitative features, maxima occurring at the close to same
grating height. Looking more closely at the result for H-polarization
in the left panel, we see that the absorption is somewhat lower for
the 5-step grating, while in all other cases the 5-step grating leads
to somewhat higher values.

We also abserve a qualitative difference between the results for the
two polarization: While in both cases E-polarization does not decrease
for larger grating height, H-polarization show a distinct maximum.
For the silver grating, this maximum occurs already at around 50nm
height, while for the polycarbonate (PC) grating it is around 80nm.
We have seen in the case of Fig. \ref{fig:Spectrum-in-H-polarisation square Ag-Si grating}
that the absorption in a silver-silicon grating becomes very weak,
while the absorption in the metal becomes very large. This is consistent
with the observation of a decreasing absorption as the grating exceeds
a certain height. We have also tested this structure replacing silver
by aluminum. The light-trapping effciency is only very little impacted
by the small additional absorption in aluminum. The choice of PC is
motivated by this consideration, but also because a planar substrate
with a PC coating that is prestructure could be used in a straightforward
way, even for mass production. How the contacts could be made in such
a device remains an experimental problem.

\subsection{Plasmons at Material Interfaces}

In reference \citep{Sheng:1982ys}, it was mentioned that complex
conjugate eigenvalues can occur in gratings with real, negative permittivities
for H polarization, and this non-obvious property was rediscovered
in references \citep{Foresti:2006uq,Sturman:2007zk}. Furthermore,
in weakly absorbing materials, these eigenvalues are not exactly conjugate
anymore, but they still occur pairwise. To obtain an accurate calculation,
both eigenvalues must either be included or omitted. However, for
the type of metallic structures studied in this paper, degenerate
eigenmodes can occur as well. It is difficult to obtain these eigenvalues
correctly with the modal method based on the transcendental equations,
because these eigenmodes have nearly exactly degenerate eigenvalues,
and are difficult to locate and especially find them both.

\begin{figure}
\includegraphics[width=5.5cm]{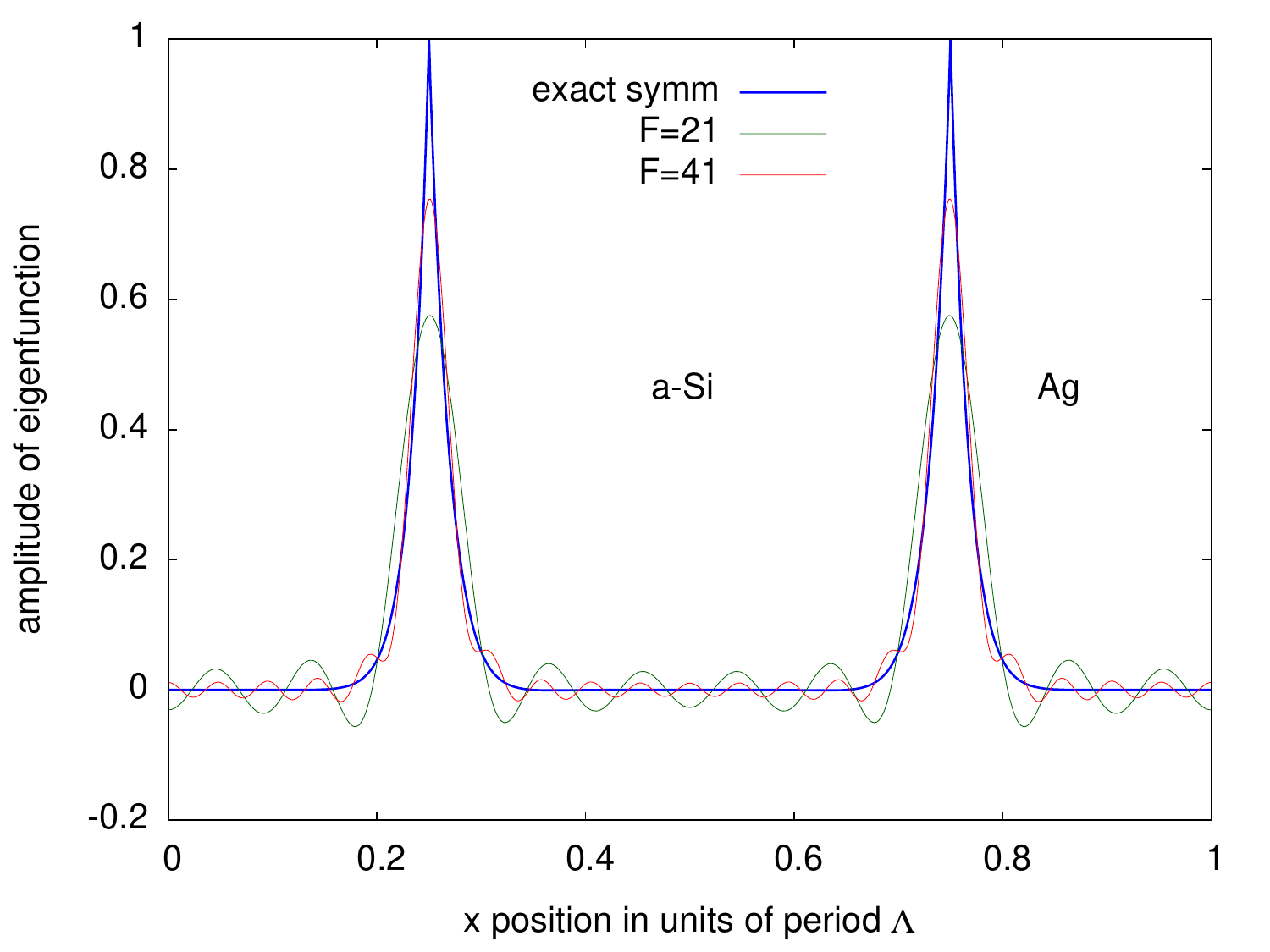}\includegraphics[width=5.5cm]{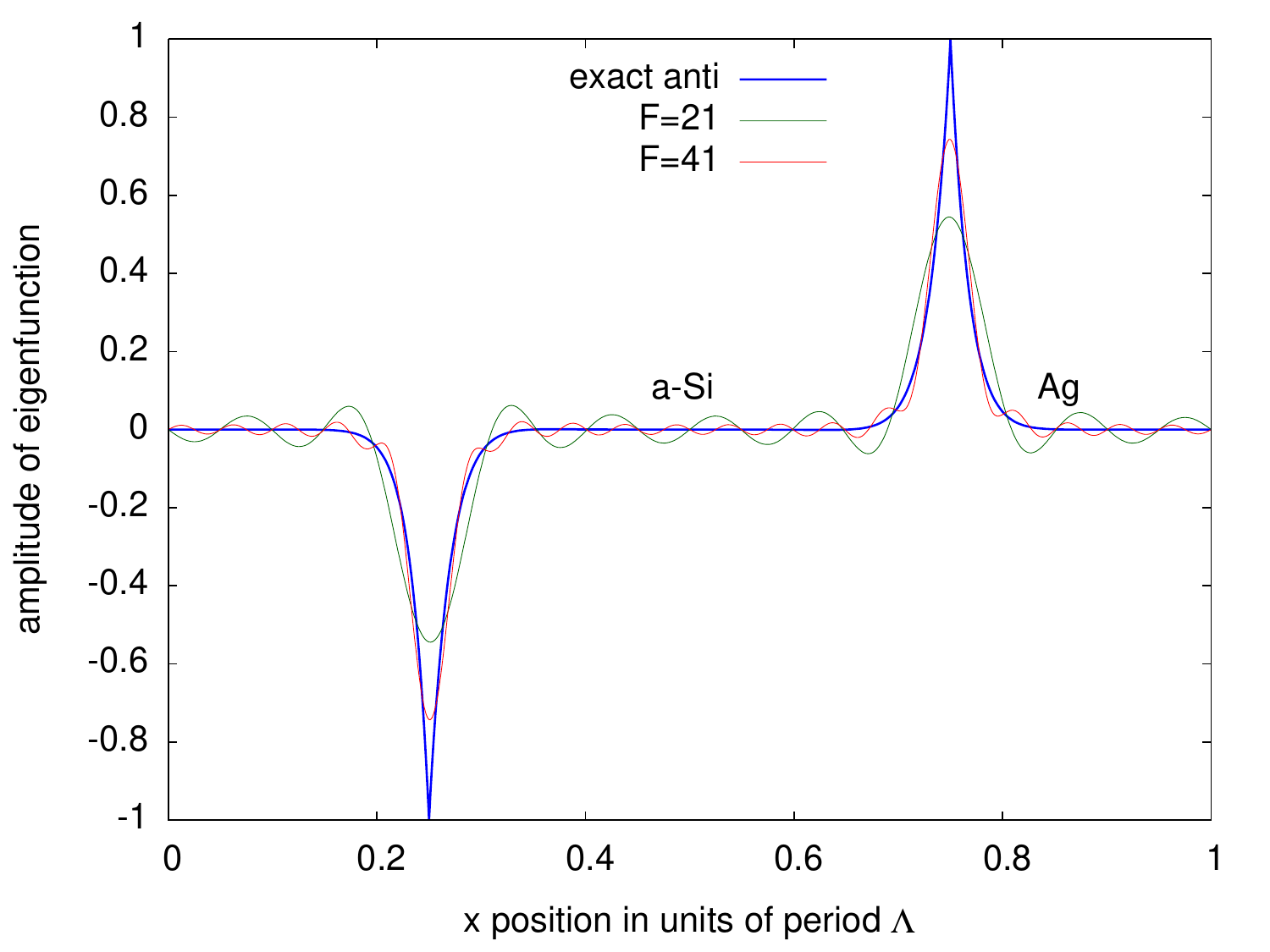}

\protect\caption{Symmetric and antisymmetric eigenfunctions for degenerate eigenvalues
in H-polarization at a wavelength $\lambda=670$nm and normal incidence.
The grating has period 511nm and consists of amorphous silicon and
silver of equal width. For comparison, we also show their Fourier
approximation with Fourier order $F=$21 and 41. These are obtained
from the exact eigenfunctions, not by solving within a Fourier modal
method.\label{fig:Eigenfunctions-for-degenerate-1}}
\end{figure}

Using the polynomial method, they are quite easy to find. But, the
fact that they occur in nearly degenerate pairs was initially a puzzle
and unexpected. Only after testing these eigenvalues and eigenfunctions
by looking at their convergence properties as the Legendre basis becomes
very large, have we been convinced that they are no mere artefact
of the polynomial method. And if the eigenvalues are known accurately,
one can confirm that they are also solutions of the transcendental
equation \citep{Botten:1981kq,Botten:1981tp,Sheng:1982ys}. These
eigenfunctions display the character of plasmons. That they are nearly
degenerate and for some wavelength numerically exactly degenerate
is the result of their exponential decay away from the interface.
If the two interfaces are sufficiently far apart, the tunnel splitting
(in the language of quantum mechanics) becomes exponentially small.
At zero angle of incidence, they can be classified by symmetry. As
the geometry of a square wave grating with only two materials has
axes of reflection symmetry, the eigenfunctions for perpendicular
incidence can be eigenfunctions of the parity operator. If however
the degeneracy is numerically exact, generally we will see superpositions
of the two eigenvectors with undefined parity. In our example depicted
in Fig. \ref{fig:Eigenfunctions-for-degenerate-1}, the square wave
grating consists of silver and amorphous silicon with a period $\Lambda=$511nm
and both materials with equal width. The wavelength is 670nm. 

These plasmonic eigenfunctions appear as a result of the boundary
conditions at the interface. In other words, the condition of $\frac{1}{\epsilon_{1}}\partial_{x}H_{y,1}=\frac{1}{\epsilon_{2}}\partial_{x}H_{y,2}$
enforces a sharp cusp in the magnetic field $H_{y}$at the interface
for H polarization, i.e. if the permittivities in different domains
have the opposite sign. They cannot appear for E polarization, where
the derivative of the field has to fulfill $\partial_{x}E_{y,1}=\partial_{x}E_{y,2}$.

Figure \ref{fig:Eigenfunctions-for-degenerate-1} shows these eigenfunctions
as well their Fourier approximations. The Fourier approximation was
calculated as a series expansion of the eigenfunction as calculated
from the corresponding numerically exact eigenfunction. The maximum
value of the Fourier approximated eigenfunction clearly remains below
the numerically exact maximum, and this Gibbs phenomenon leads to
a significant underestimation of the maximum field value of the $H_{y}$
field. More importantly, at the cusp, the derivative $\frac{1}{\epsilon}\partial_{x}H_{y}=E_{z}$
of the exact eigenfunction is very large, whereas for any Fourier
expansion based method the electric field component for $E_{z}\sim\frac{1}{\epsilon}\partial_{x}H_{y}$
vanishes at the interface, which is clearly wrong. Correspondingly,
the current at the interface will vanish as well and thus the absorption
near the surface will be significantly underestimated.

\begin{figure}[h]
\includegraphics[clip,width=6cm]{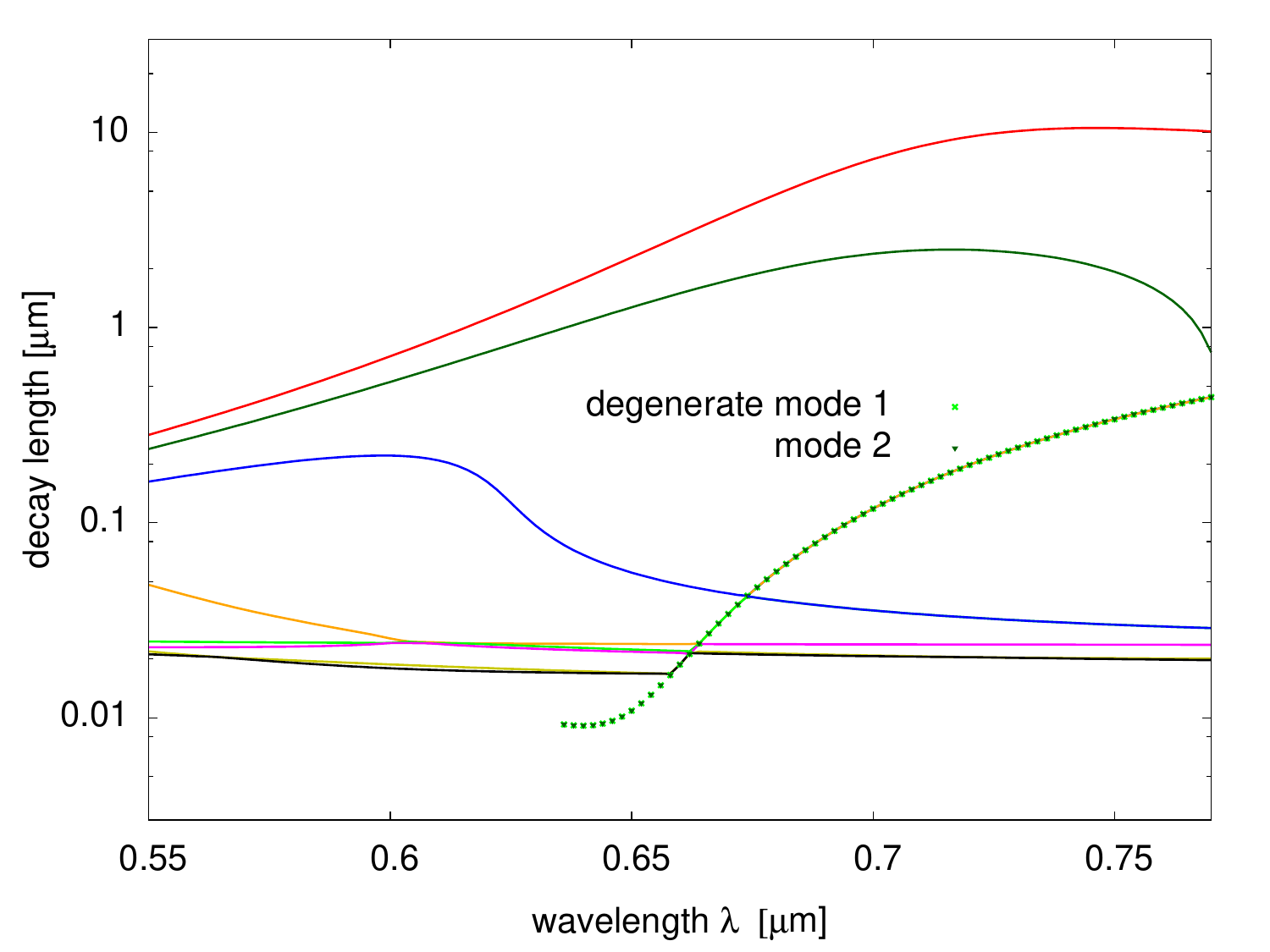}

\protect\caption{Spectrum of the few modes with largest decay length inside the grating
of period 511nm and aluminum and amorphous silicon of equal width.
The two degenerate modes become very important as the wavelength increases
beyond about 680nm.\label{fig:SpectrumSplitting}}
\end{figure}

Figure \ref{fig:SpectrumSplitting} shows the mode spectrum of the
modes with longest decay length. The degenerate modes start with very
short decay length of less than 10nm around a wavelength of 640nm.
The decay length is is a measure how far a mode can couple in the
z-direction. With increasing wavelength above 680nm, the two degenerate
plasmonic modes become propagating within the grating layer as its
thickness is at most around 100nm. In the spectral range from 680
to 770nm they are the third and fourth most important modes carrying
the energy flux. They do contribute to the absorption in the metal
seen in Fig.6 and 8, which gets very large in H-polarization. Clearly
their omission from a modal calculation would give wrong results.
As they are very difficult to locate in the complex plane, the calculation
of these eigenvalues and eigenfunctions is extremely difficult if
we follow the treatment of refs. \citep{Botten:1981kq,Botten:1981tp,Sheng:1982ys}.
By contrast, using the method of polynomial expansion of ref. \citep{Morf:1995fi},
they are found directly and without missing modes ordered according
to their number of nodes and up to a maximum number of nodes with
a required precision which determines the necessary size of the polynomial
basis.

2010 Gundu and Mafi reported an apparent non-convergence for the spectra
of lossless metal gratings as calculated with modal methods \citep{Gundu:2010fk},
and reported a method to reduce the problem in \citep{Gundu:2010uq}.
Li and Granet revealed the role of edge and hyper-singularities for
such gratings in \citep{Li:2011vn,Li:2012yq,Li:2014vn}, and Mei,
Liu and Zhong proposed a method that gradually varied the index of
refraction over the boundary region in \citep{Mei:2014nx}. While
the silver back-reflector in solar cells is not lossless, it is highly
conducting nevertheless, and remains a delicate calculation, given
that the skin depth of the metal for the studied wavelength range
of $600-800$ nm is on the order of 10nm. This skin depth of course
also sets a minimum resolution requirement for numerical solution.

\subsection{\textcolor{black}{Depth Dependence}}

\textcolor{black}{In this section, we would like to give a short overview
of the influence of several grating parameters on the absorption in
the amorphous silicon part of the pseudo-sine gratings described above.
Here and in the following, we wish to point out, that when the grating
depth is varied, the total amount of amorphous silicon is always kept
constant. This is done by reducing the thickness of the homogenous
amorphous silicon layer. The reason for this choice is that we do
not want to mix the effects of absorption by light-trapping with those
resulting from modifying the total amount of silicon present in the
structure.}

\subsubsection{\textcolor{black}{Amorphous Silicon Thickness and Grating Depth}}

\textcolor{black}{The first grating parameter that can be investigated
is the silicon thickness. For that, consider the first case of a silver
grating case, where the total silicon thickness is gradually reduced.
The left panel of figure \ref{fig:Asithickness-1} shows that for
E polarization, the a-Si thicknesses $170-200$nm have very little
influence on the integrated absorption curve for grating depths between
$20$ nm and $50$ nm. For H polarization, the right panel of figure
\ref{fig:Asithickness-1} shows that the integrated absorption decreases
for all grating depths with decreasing a-Si layer thickness. }

\noindent \begin{center}
\textcolor{black}{}
\begin{figure}[h]
\textcolor{black}{\includegraphics[width=0.5\columnwidth]{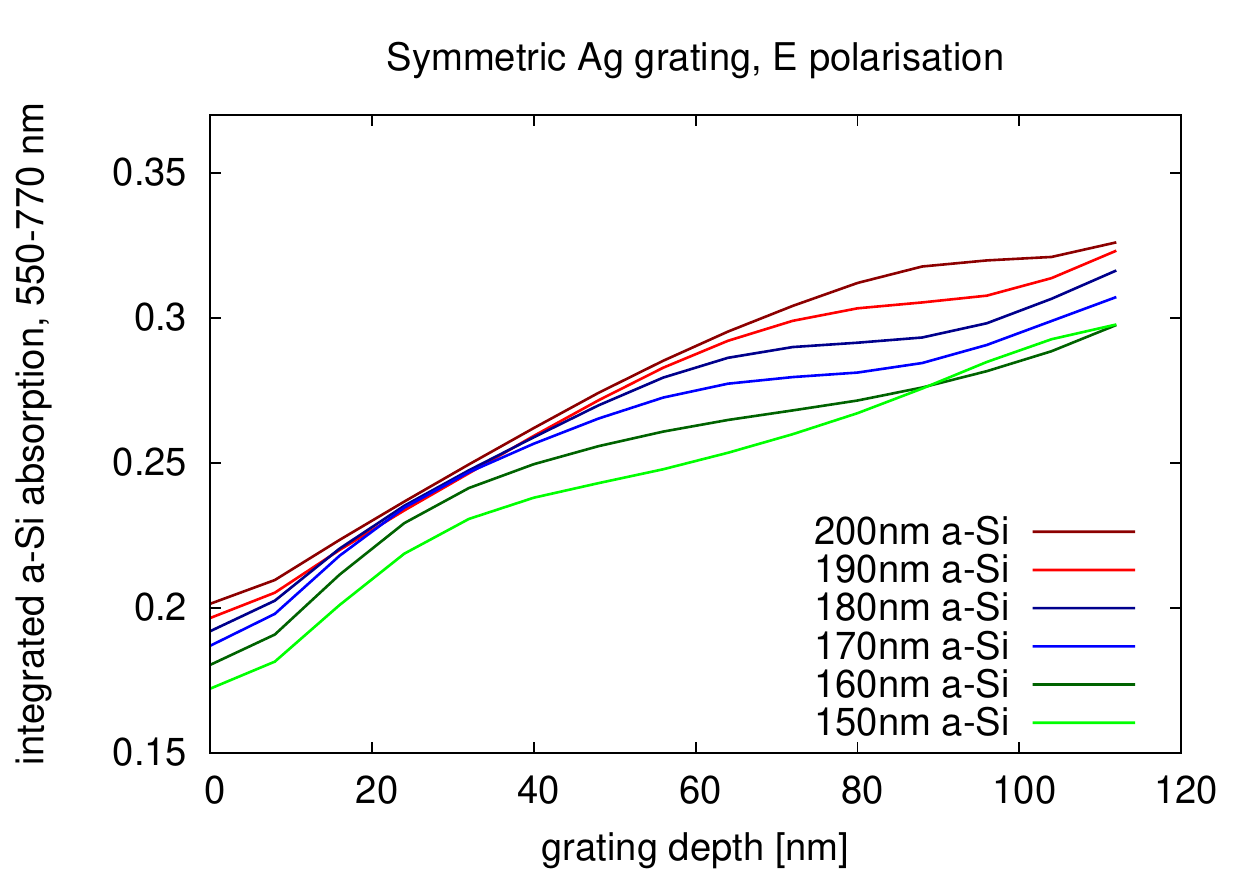}\includegraphics[width=0.5\columnwidth]{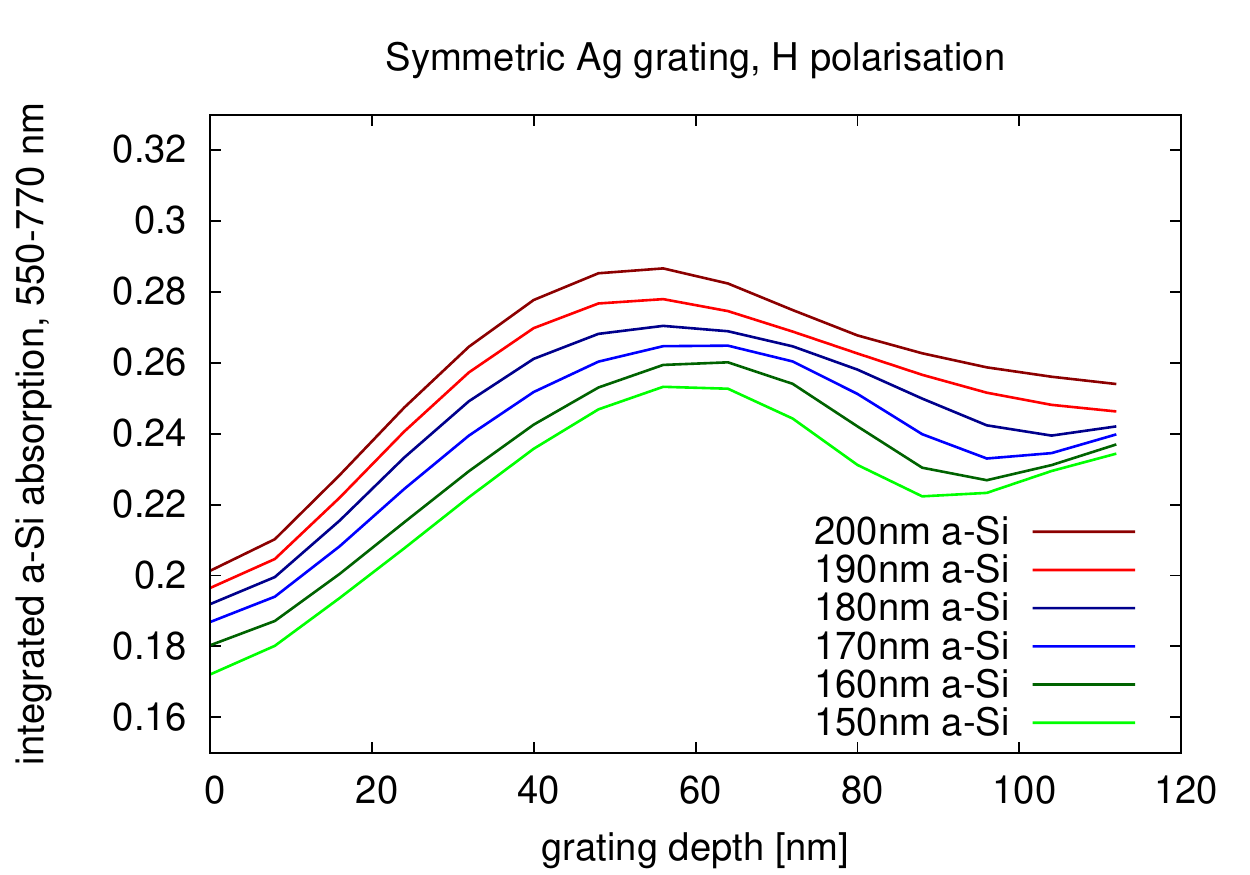}\protect\caption{The integrated absorption 550-770 nm for both polarizations as a function
of grating depth, for a structure of type $P^{\left(9\right)}\left(\mbox{Ag|a-Si}\right)|\mbox{AR}^{*}$.
Left Panel: E polarization. Right panel: H polarization.\label{fig:Asithickness-1}}
}
\end{figure}

\par\end{center}

\begin{figure}[h]
\includegraphics[width=5cm]{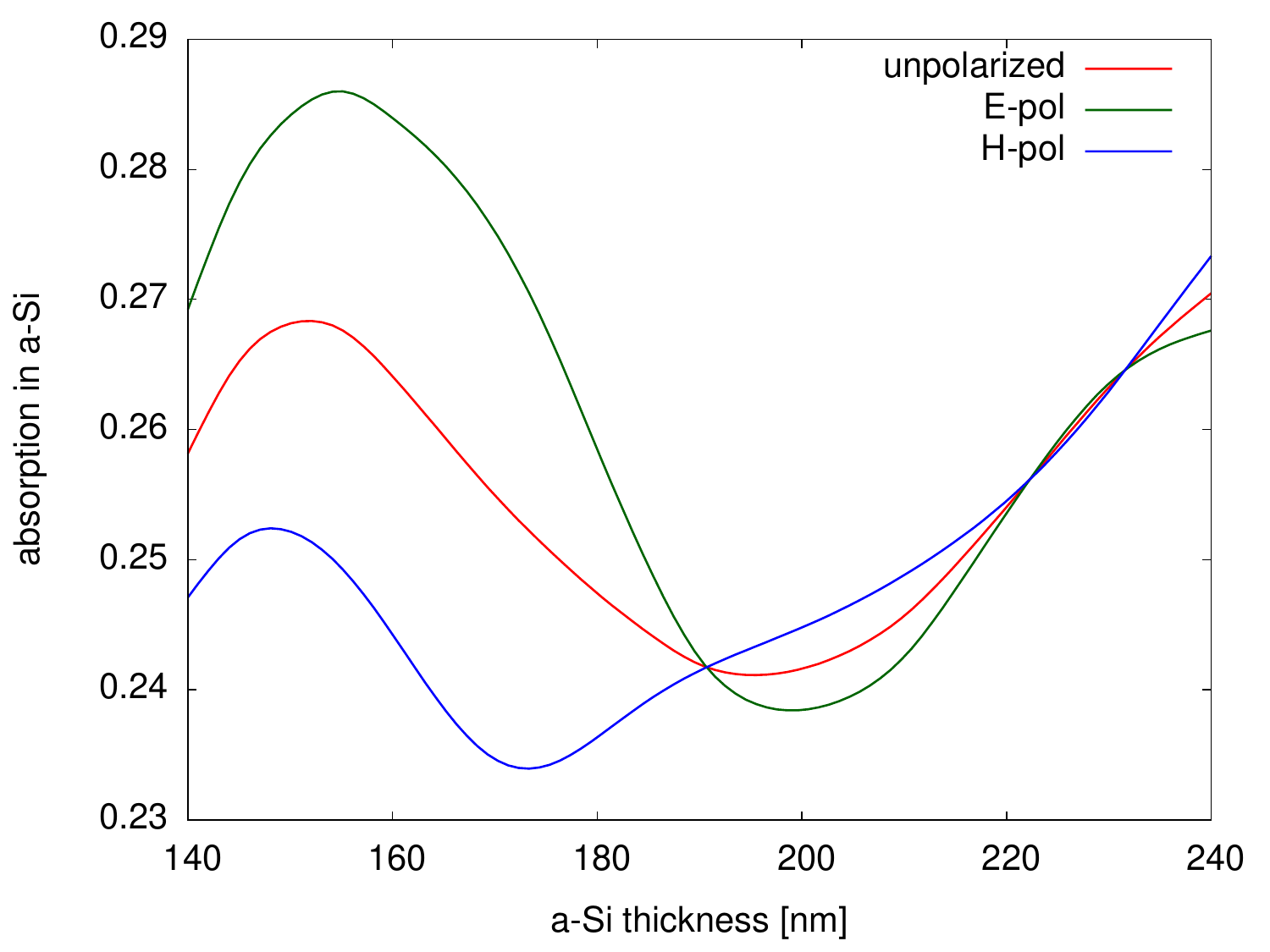}\protect\caption{Integrated absorption 550-770 nm, weighted with AM1.5, for a rectangular
grating consisting of $\mbox{Ag|ZnO|a-Si|ITO},$ where the a-Si and
ITO layers are less deep than the grating (60nm) and for a varying
silicon thickness.\label{fig:integrated-absorptionASI}}
\end{figure}

Figures \ref{fig:integrated-absorptionASI} show that contrary to
intuition, the integrated absorption contributions for 550-770nm for
a simple square-wave grating are not just increasing for increasing
a-Si layer thickness, but rather that there is a pronounced maximum
at about 150nm, i.e. substantially below the 200nm thickness that
is often the preferred choice in experiments. 

\subsubsection{\textcolor{black}{Polycarbonate Thickness}}

\textcolor{black}{For the second case of a polycarbonate grating,
the homogeneous polycarbonate layer thickness can be varied, as shown
in figure\ref{fig:PCthickness}. The left panel of figure \ref{fig:PCthickness}
shows that for E polarization there is an almost linear relationship
between the grating depth and the a-Si absorption, and the spread
between the PC layer grating thickness sets in gradually. For the
H polarization shown in the right panel of \ref{fig:PCthickness}
a thicker PC layer increases the absorption only very little for grating
depths below 60 nm, whereas the absorption increases significantly
with the $80$ nm and $100$ nm layers for gratings with a depth greater
than $80$ nm. }

\begin{center}
\textcolor{black}{}
\begin{figure}
\textcolor{black}{\includegraphics[width=0.5\columnwidth]{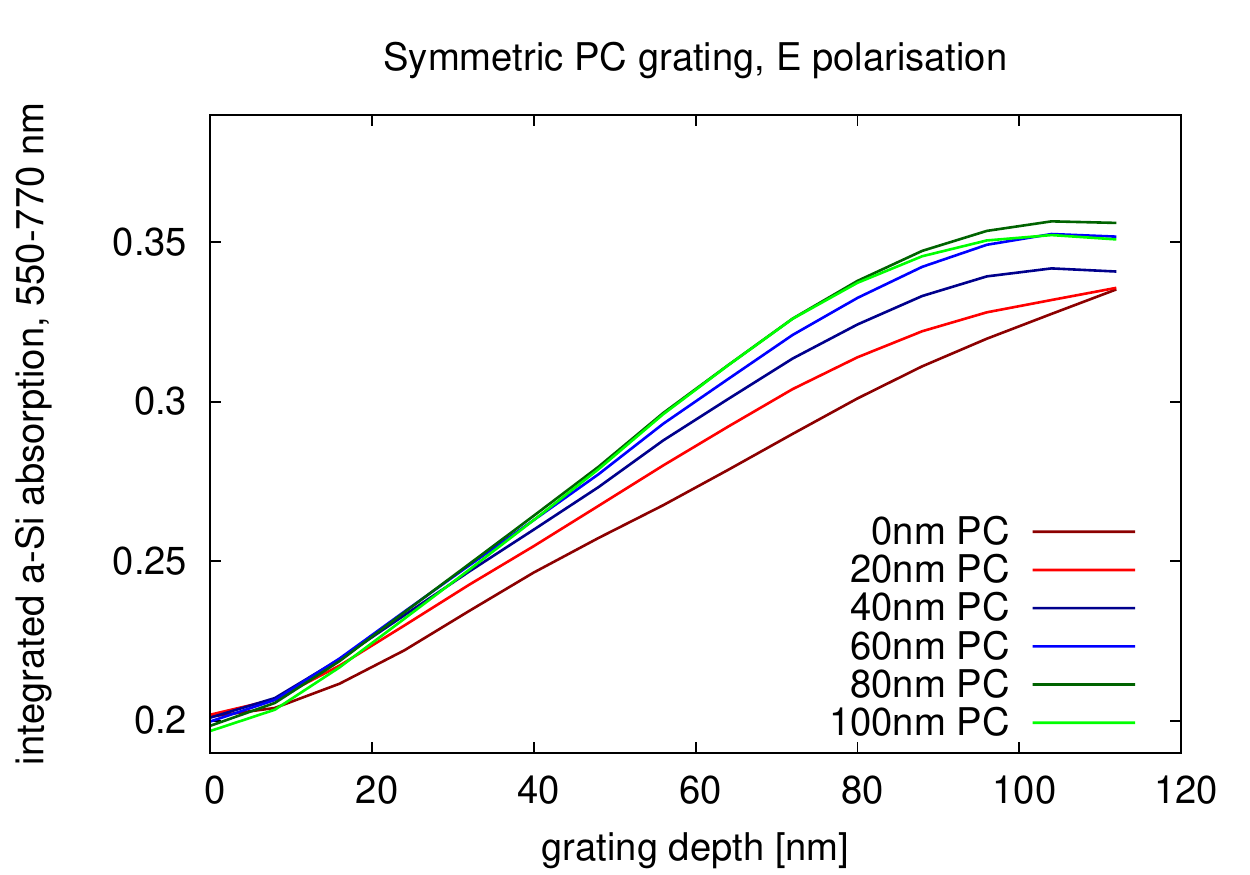}\includegraphics[width=0.5\columnwidth]{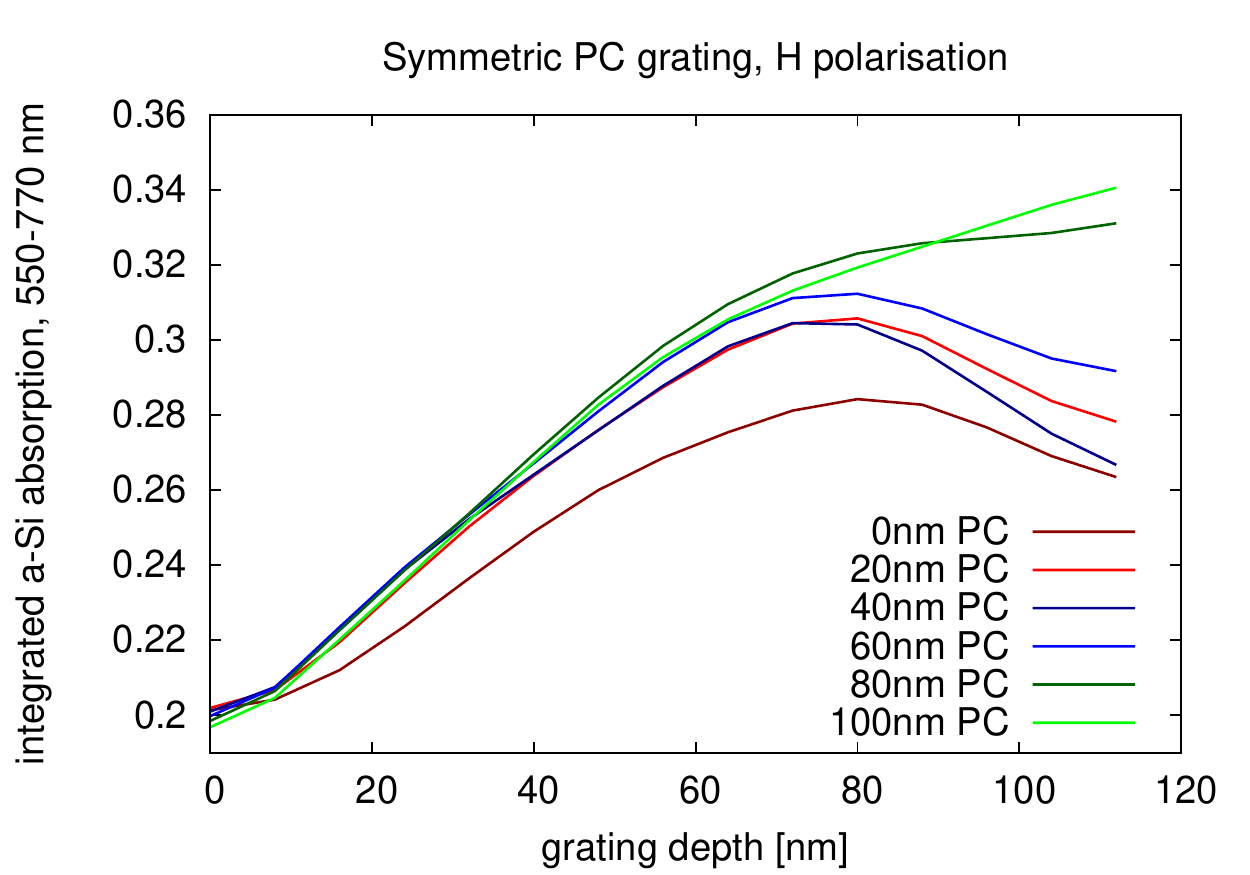}\protect\caption{The integrated absorption in a-Si for both polarizations as a function
of grating depth, for a grating of type $\mbox{PC|}P^{\left(9\right)}\left(\mbox{PC|}\mbox{a-Si}\right)|\mbox{AR}^{*}$
. Left Panel: E polarization. Right panel: H polarization.\label{fig:PCthickness}}
}
\end{figure}

\par\end{center}

\subsubsection{Dependence on thickness of ZnO and ITO}

\begin{figure}[h]
\includegraphics[width=5cm]{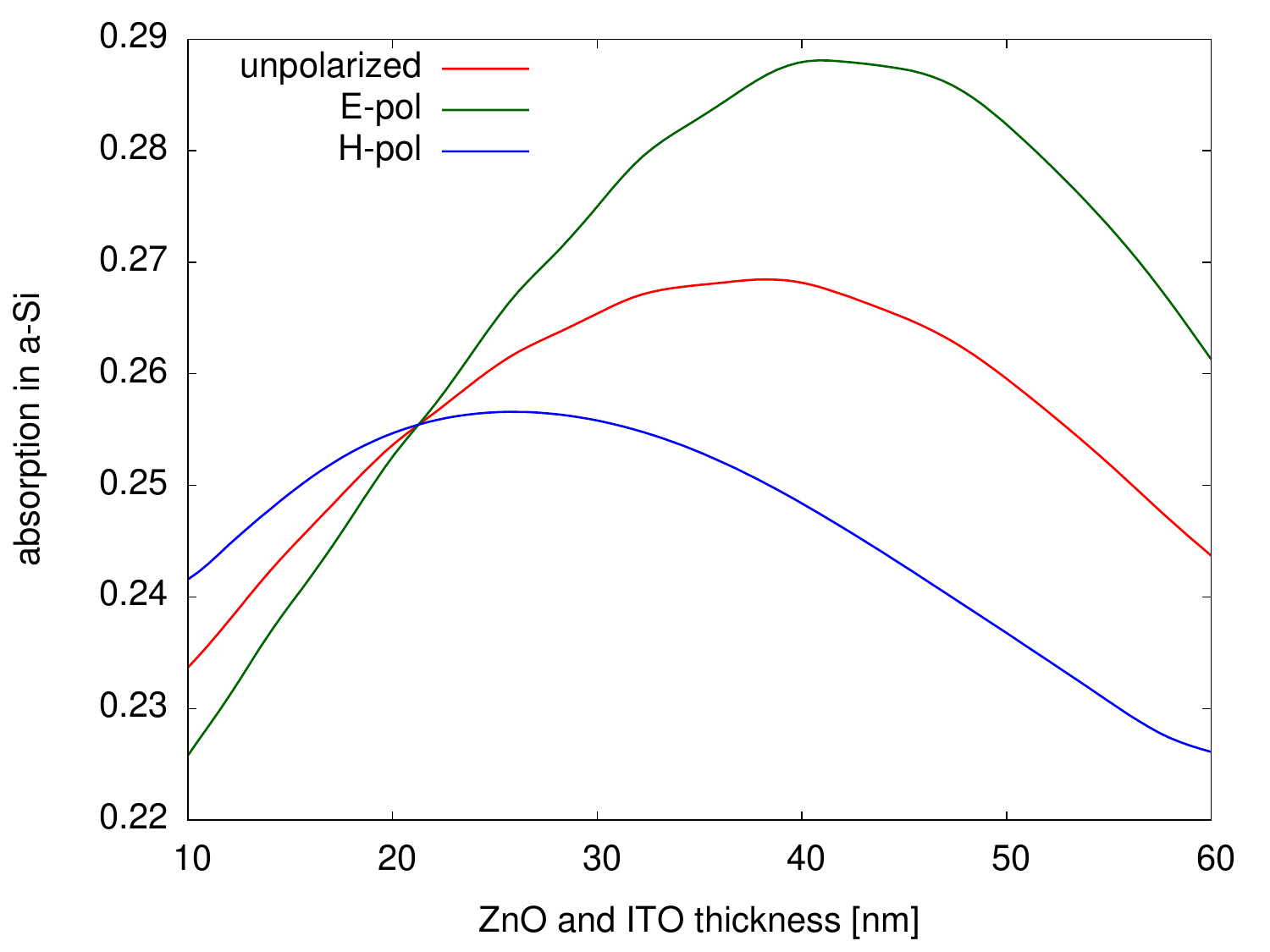}\protect\caption{Integrated absorption 550-770 nm, weighted with AM1.5 spectrum, as
function of ITO and ZNO layer thickness, for a rectangular grating
consisting of $\mbox{Ag|ZnO|a-Si|ITO},$ where the a-Si and ITO layers
are less deep than the grating.\label{fig:integrated-absorptionITOZNO}}
\end{figure}

In figure \ref{fig:integrated-absorptionITOZNO}, we depict the behavior
of the absorption as function of thickness of the ITO and ZnO layers.
Clearly, the optimal values are quite different for the two polarization.

\textcolor{black}{}

\subsection{\textcolor{black}{Period Dependence}}

\textcolor{black}{The influence of the grating period on the absorption
is another parameter that can be studied. For that, consider the case
of a ZnO grating on an aluminum substrate with 200nm amorphous silicon
and an ideal antireflection layer on top of the silicon. The calculations
for figures \ref{fig:perioddepDepth} and \ref{fig:perioddepZNO}
were performed with 31 numerically exact eigenmodes. For H polarization,
the right panel of figure \ref{fig:perioddepDepth} shows that the
optimal period for maximal absorption in the amorphous silicon does
not change, and that in the region of the optimal period the deepest
gratings perform best. For E polarization, there is a material specific
absorption optimum below the deepest structure with the largest period.
This local optimum moves from about $0.44\,\textrm{\ensuremath{\mu}m}$
for a grating depth of $64$ nm to about $0.5\,\textrm{\ensuremath{\mu}m}$
for a grating depth of $112$ nm. }

\begin{center}
\textcolor{black}{}
\begin{figure}
\textcolor{black}{\includegraphics[width=0.5\columnwidth]{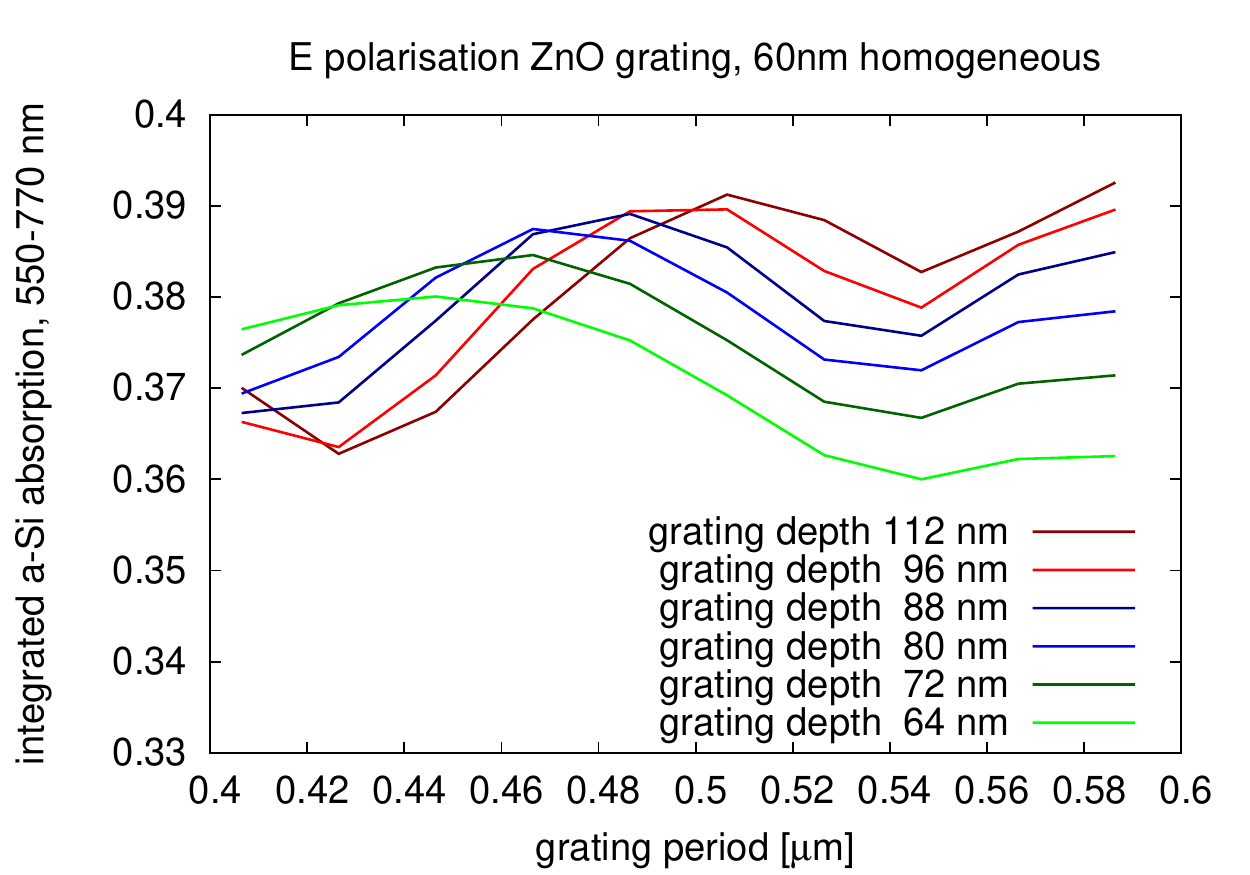}\includegraphics[width=0.5\columnwidth]{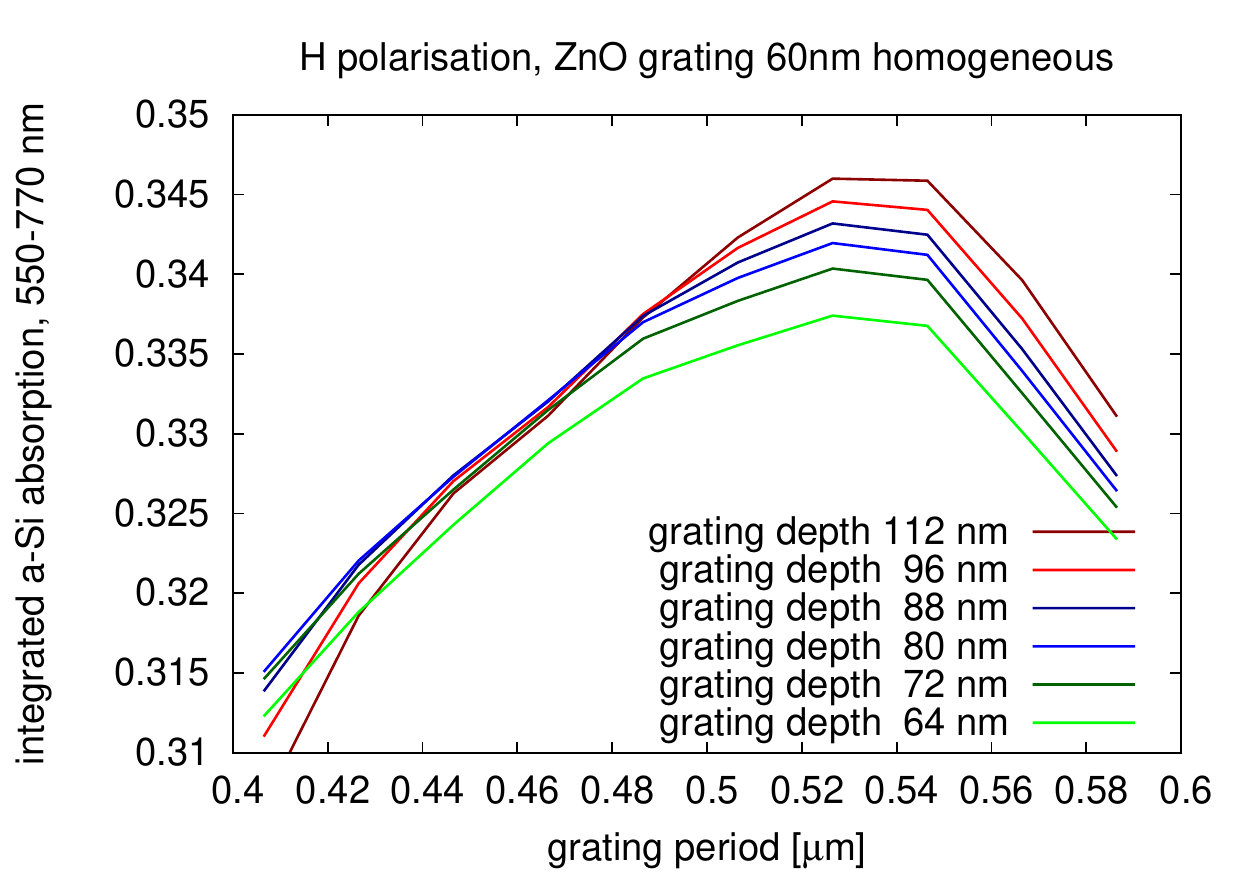}\protect\caption{The integrated absorption for both polarizations as a function of
grating period, for a structure of type $\mbox{Al|ZnO|}P^{\left(7\right)}\left(\mbox{ZnO|a-Si}\right)|\mbox{AR}^{*}$,
where the amplitude of $P^{\left(7\right)}$ is varied. Left panel:
E polarization. Right panel: H polarization.\label{fig:perioddepDepth}}
}
\end{figure}

\par\end{center}

\textcolor{black}{Figure \ref{fig:perioddepZNO} shows the period
dependence of the integrated absorption for various thicknesses of
the homogeneous ZnO layer below the ZnO grating. Both panels show
that the ZnO layer thickness has a much stronger influence on the
absorption than the period. The left panel shows that the optimum
homogeneous ZnO layer thickness should be around $40$ nm for E polarization,
and the right panel shows that for H polarization, the optimum thickness
should be around $80$ nm.}

\begin{center}
\textcolor{black}{}
\begin{figure}
\textcolor{black}{\includegraphics[width=0.5\columnwidth]{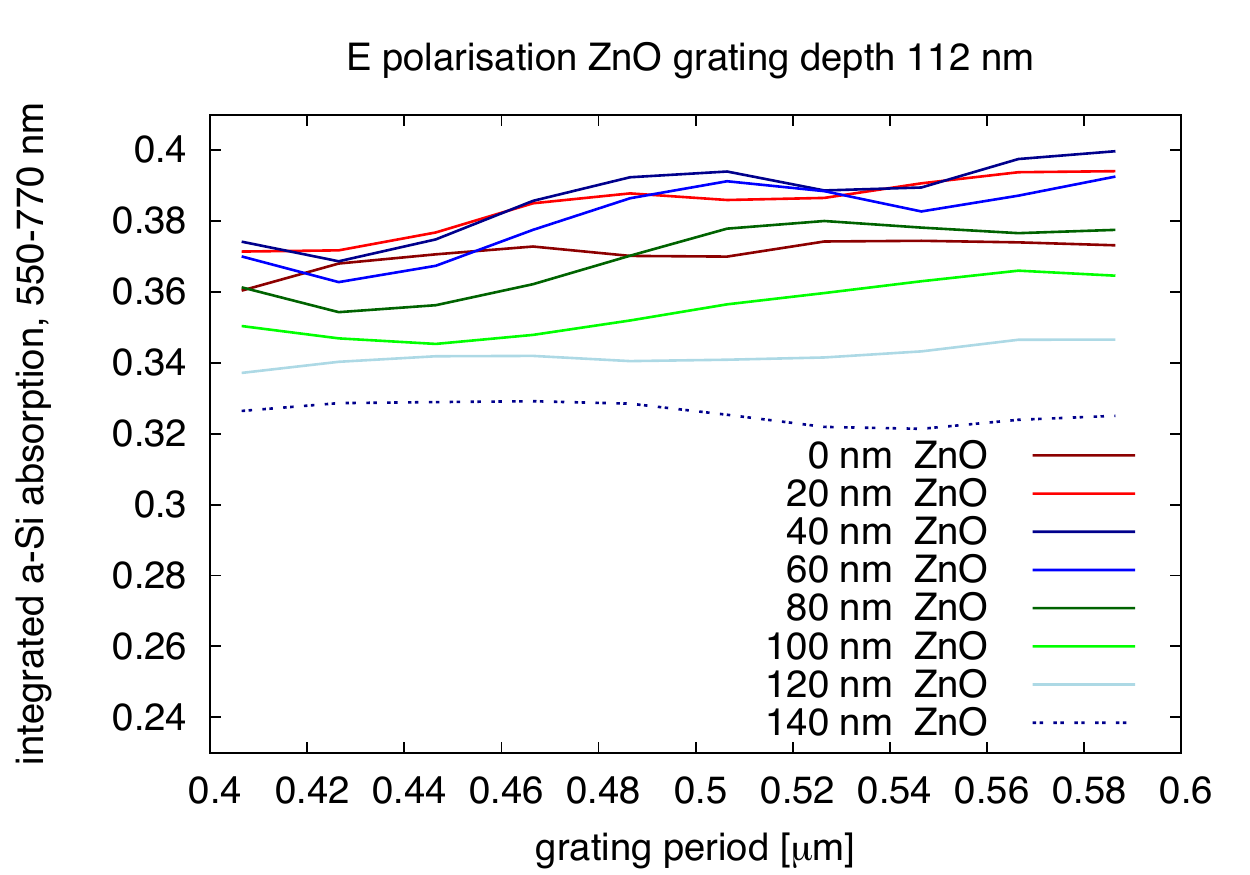}\includegraphics[width=0.5\columnwidth]{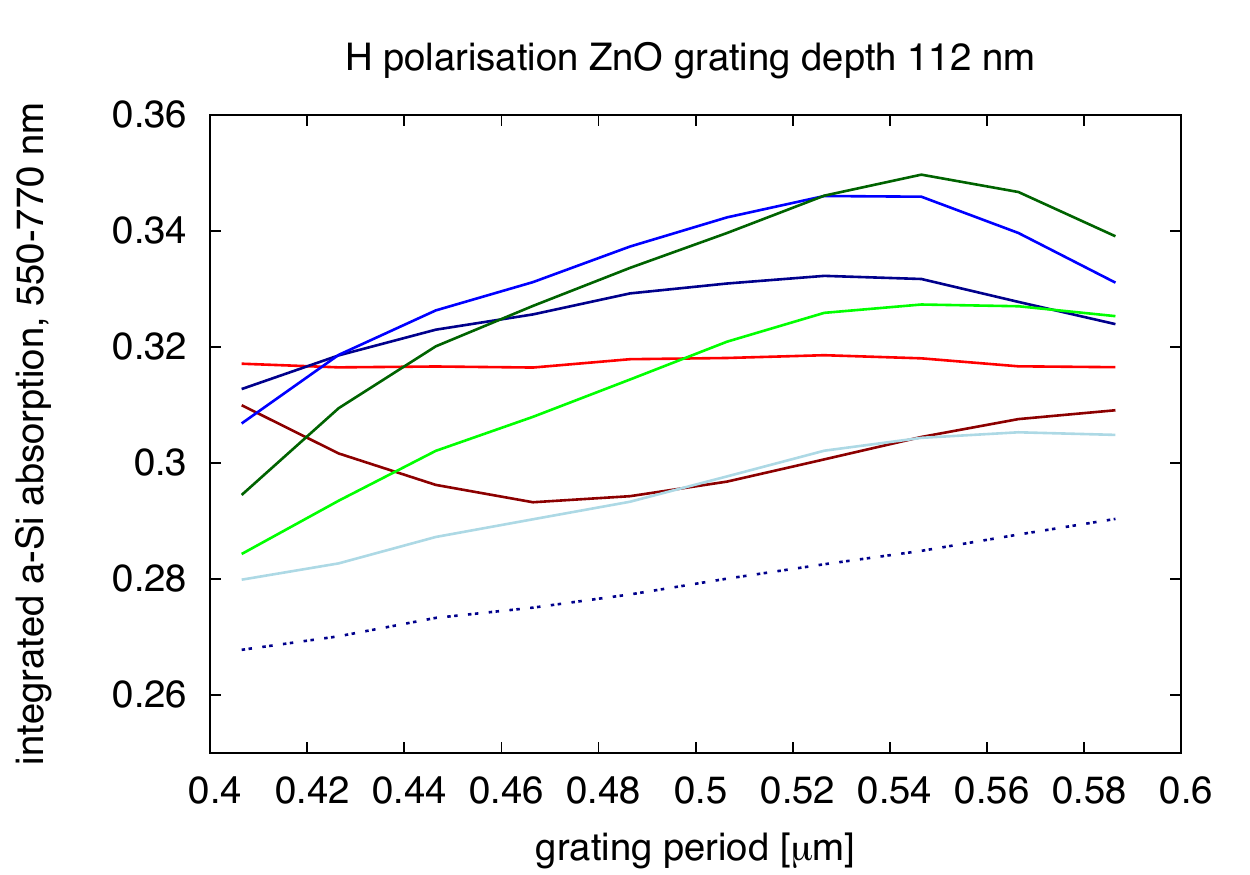}\protect\caption{The integrated absorption for both polarizations as a function of
grating period, for a structure of type $\mbox{Al|ZnO|}P^{\left(7\right)}\left(\mbox{ZnO|a-Si}\right)|\mbox{AR}^{*}$,
where the homogeneous $\mbox{ZnO}$ layer is varied. Left panel: E
polarization. Right panel: H polarization.\label{fig:perioddepZNO}}
}
\end{figure}

\par\end{center}

\subsection{\textcolor{black}{Larger Periods}}

\textcolor{black}{So far, all investigated gratings have been using
periods in the $400-600$ nm period range. This section will show
that larger periods will not lead to an increase in absorption in
the $550-770$ nm wavelength range. For this exploration, the example
grating structure consisted of a flat silver back reflector, covered
by a 7-step pseudo-sine polycarbonate grating which is covered with
a silicon layer that is on average $200$ nm thick and flat on top.
The grating is illuminated perpendicularly. The choice of a flat reflector
at the bottom should be at the upper end of the attainable absorption
for the amorphous silicon, given that it was shown section IV.A.2
that a substantial amount of absorption takes place in the metallic
grating. }

\begin{center}
\textcolor{black}{}
\begin{figure}
\textcolor{black}{\includegraphics[width=0.5\columnwidth]{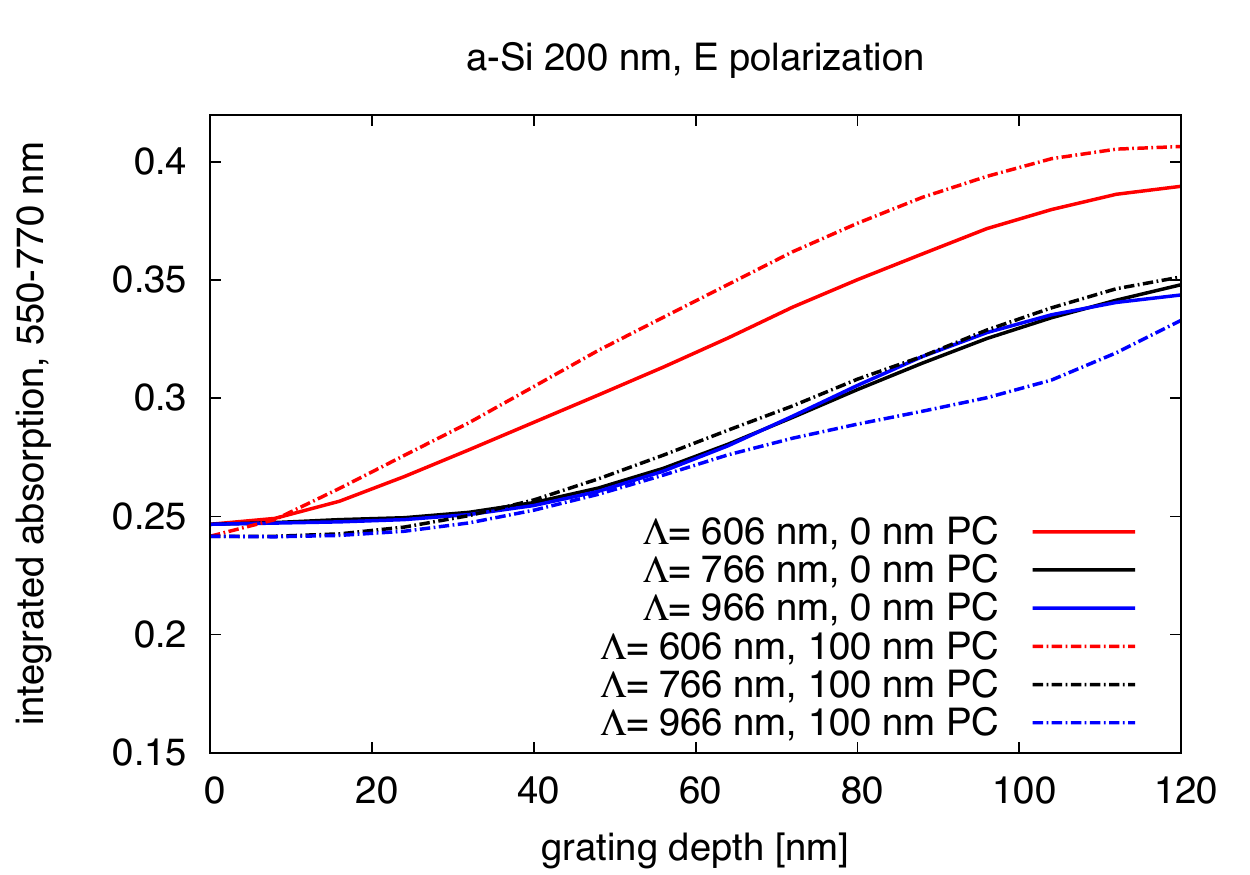}\includegraphics[width=0.5\columnwidth]{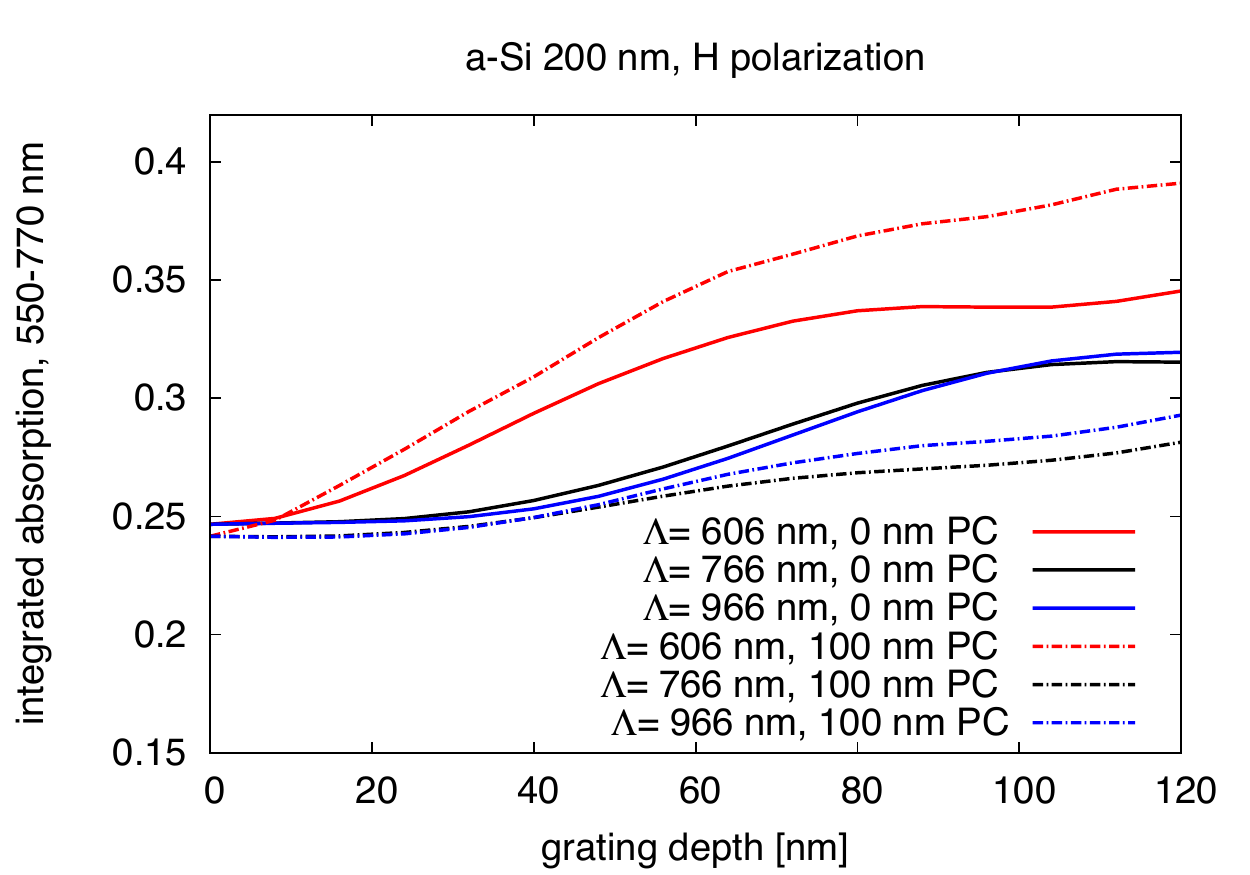}}

\textcolor{black}{\protect\caption{The depth dependence of the integrated absorption for both polarizations
for a structure of type $\mbox{Ag|PC|}P^{\left(7\right)}\left(\mbox{PC|a-Si}\right)|\mbox{AR}^{*}$,
where the PC layer is either 0 nm or 100 nm thick.\label{fig:the-depth-dependence}}
}
\end{figure}

\par\end{center}

\textcolor{black}{In figure \ref{fig:the-depth-dependence}, the depth
dependence of the integrated absorption as a fraction of the AM1.5
spectrum is shown, and in figure \ref{fig:the-period-dependence}
the period dependence is shown for select grating depths. Figure \ref{fig:the-period-dependence}
clearly shows that periods above 600nm are not helpful to increase
absorption in the chosen spectral region for both polarization, although
the effect appears to be more drastic for H polarization. Figure \ref{fig:the-depth-dependence}
shows that increasing the homogeneous polycarbonate layer between
silver and amorphous silicon does not increase absorption for the
gratings with longer periods, but it may do so at $600$ nm and below.
In all cases, deep gratings increase the absorption the most, however
for real applications, their depth would have to remain limited for
several reasons, including the manufacturing process and the need
to establish electrical contacts. In figure \ref{fig:the-period-dependence},
the homogeneous polycarbonate layer thickness is $100$ nm. The calculation
presented made use of 31 modes in each layer.}

\begin{center}
\textcolor{black}{}
\begin{figure}
\textcolor{black}{\includegraphics[width=0.5\columnwidth]{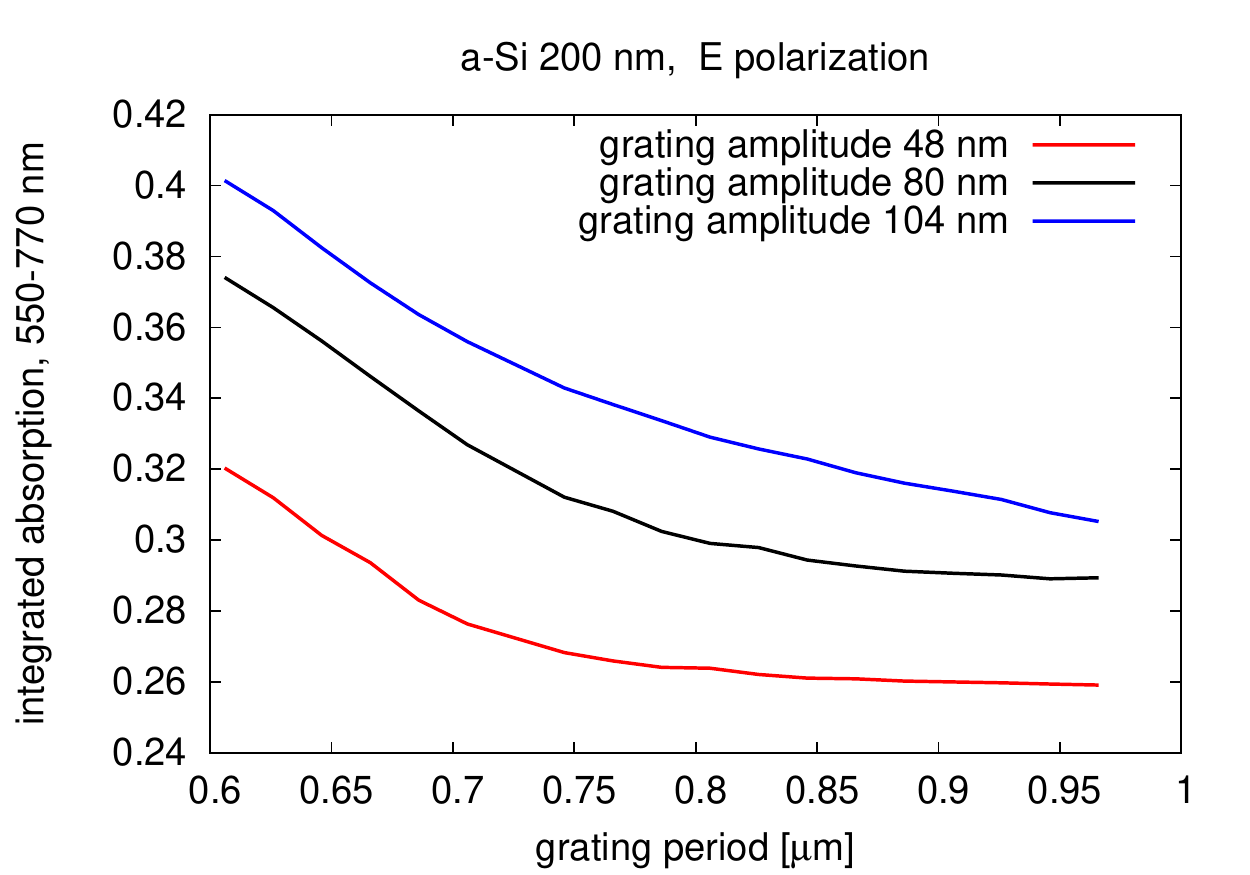}\includegraphics[width=0.5\columnwidth]{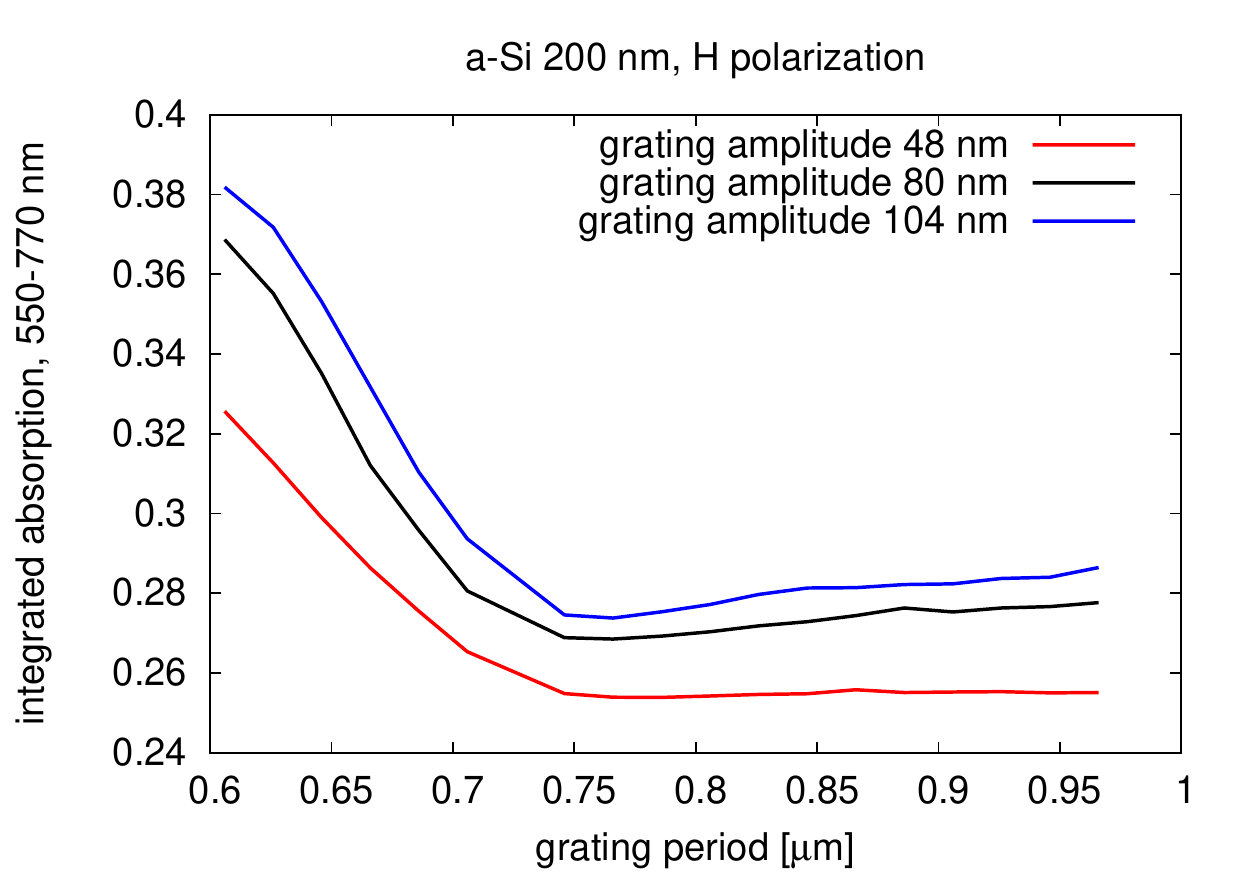}}

\textcolor{black}{\protect\caption{The period dependence of the integrated absorption for both polarizations
$\mbox{Ag|PC|}P^{\left(7\right)}\left(\mbox{PC|a-Si}\right)|\mbox{AR}^{*}$.\label{fig:the-period-dependence}}
}
\end{figure}

\par\end{center}

\subsection{Improving metallic gratings by intermediate dielectric\label{sub:Improving-metallic-gratings}}

In section \ref{sub:Energy-conservation}, we have observed the significant
absorption occurring in a (silver | a-Si) grating. Here, we examine
if the losses can be reduced by inserting a ZnO layer between silver
and a-Si, i.e. by using instead (silver | ZnO | a-Si) inside the grating.
That this can be successful is shown in Figure 19 for a symmetric
3-step pseudo-sine with period 566.1nm. The integrated absorption
for 600$<\lambda<$770nm, AM1.5-weighted, shows that the ZnO lining
of width 11nm on either side of the silver domain while reducing the
silver width by 20nm improves the absorption in H-polarization for
larger grating depth by about 16 percent. In E-polarization, the absorption
is virtually unchanged. While the absorption in E-polarization steadily
increases with increasing grating depth, in H-polarization it shows
a maximum around a depth 50nm (unlined) and around 70nm with ZnO lining
of width 11nm.

\begin{figure}

\includegraphics[clip,width=5.5cm]{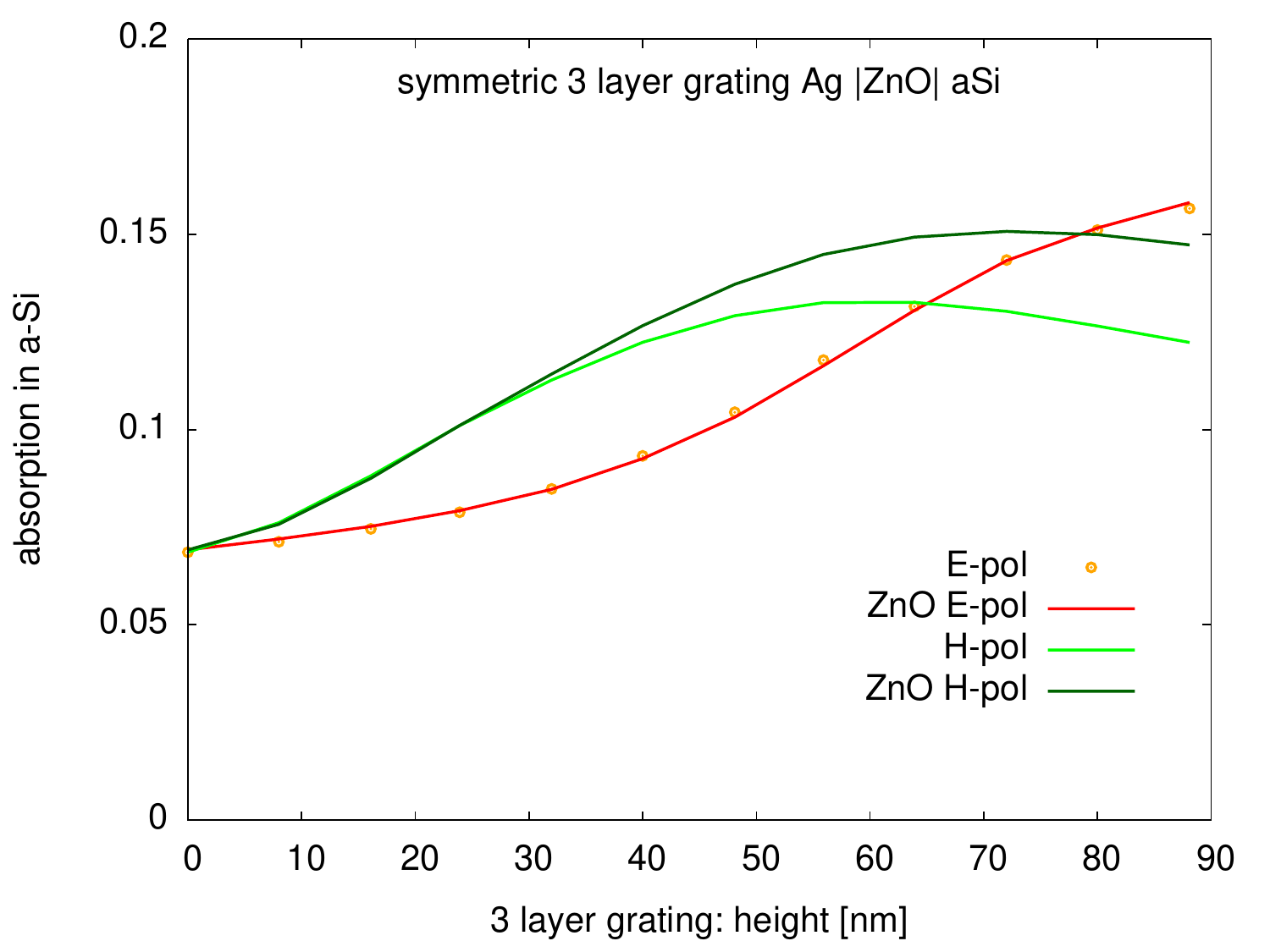}\protect\caption{Improvement of integrated absorption (600nm$<\lambda<770nm)$ for
a 3-step pseudosine grating with 566.1nm period, as a 11nm ZnO lining
is inserted between a-Si and Ag. H-polarization is improved, E-polarization
is basically unchanged. The grating structure is Ag|$P^{(3)}$(a-Si
| Ag ) | a-Si |$P^{(3)}$(ITO | a-Si ) | ITO, in which Ag is either
lined with 11nm ZnO, or not lined, as indicated in the plot labels.
The ITO layer is 70nm thick.\label{fig:Improvement-of-integrated}}

\end{figure}

\subsection{Improving absorption by asymmetric gratings}

Here, we examine the effect of introducing geometric asymmetry in
the 3-step pseudo-sine of section \ref{sub:Improving-metallic-gratings},
by shifting the the layers forming the 3-step pseudo-sine of Fig.
19 by 0, 12 and 24 percent in units of the grating period $\Lambda$=566.1nm,
respectively. Qualitatively the behavior is not different from the
symmetric grating case. Quantitatively, we observe that in the non-symmetric
structure the absorption in E-polarization is increased by about 15
percent, while for H-polarization it is increased by only about 6
percent, when compared to the corresponding symmetric results of Fig.
\ref{fig:Improvement-of-integrated}.

\begin{figure}
\includegraphics[clip,width=5.5cm]{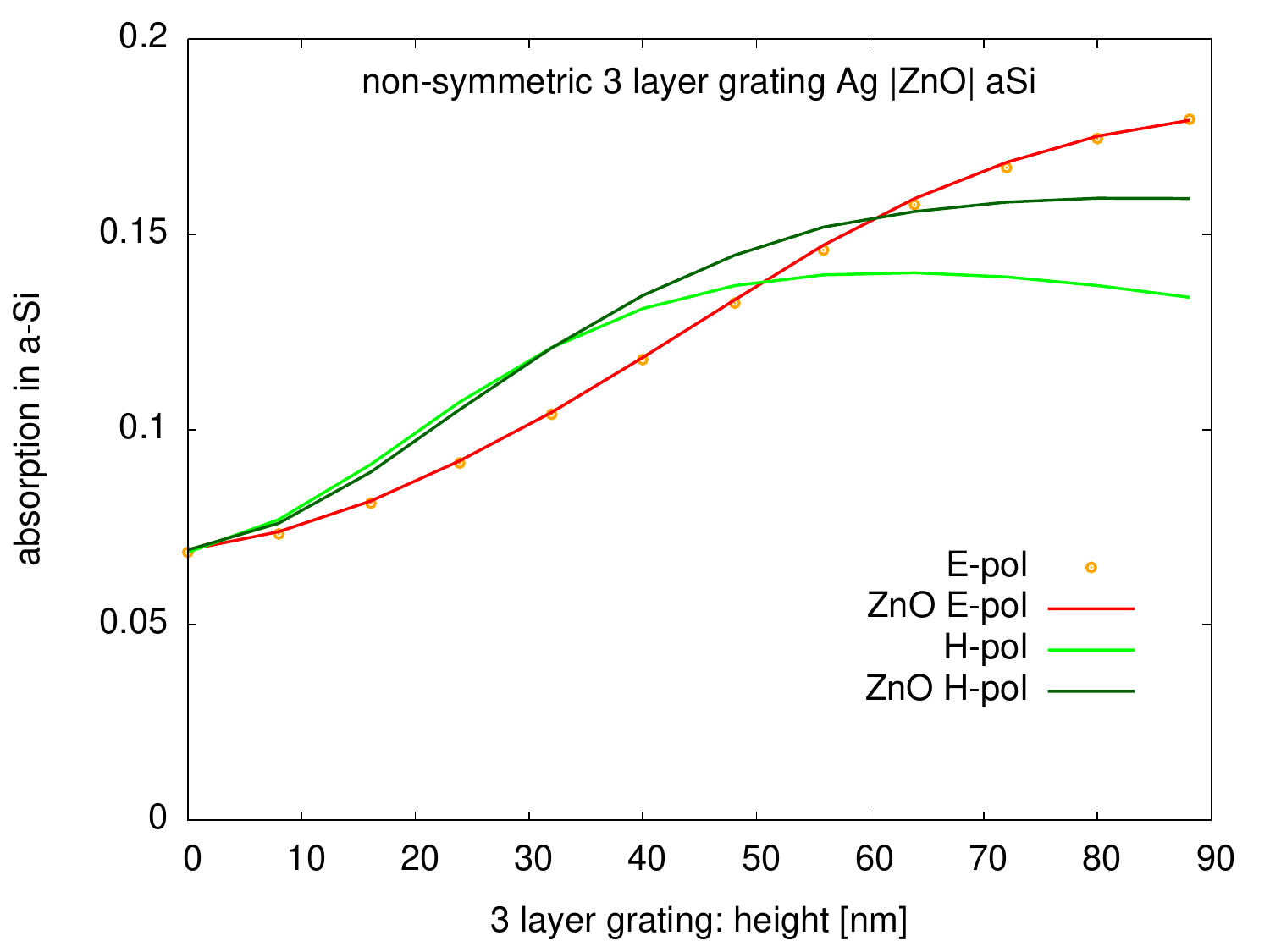}\protect\caption{Non-symmetric grating structure using same sequence of grating layers
as in Fig. \ref{fig:Improvement-of-integrated}, but with a shift
of 0, 11.5 and 23 percent for the 3 layers forming the 3-step pseudo-sine.\label{fig:Non-symmetric-grating-structure}}
\end{figure}

For completeness, we also show the spectra corresponding to the gratings
structures of Figures \ref{fig:Improvement-of-integrated} and \ref{fig:Non-symmetric-grating-structure}
as well as a comparison between ZnO-lined and unlined spectra for
the non-symmetric case. We  can see that introducing asymmetry gives
rise to additional peaks in the absorption spectrum, which in H-polarization
are only present if the metal-semiconductor interface is modified
by an intermediate ZnO layer (right panel). For E-polarization, the
effect of the ZnO lining leads to a slight shift in the position of
the peaks with no effect on their height and very little effect on
their area.

\begin{figure}
\includegraphics[clip,width=4cm]{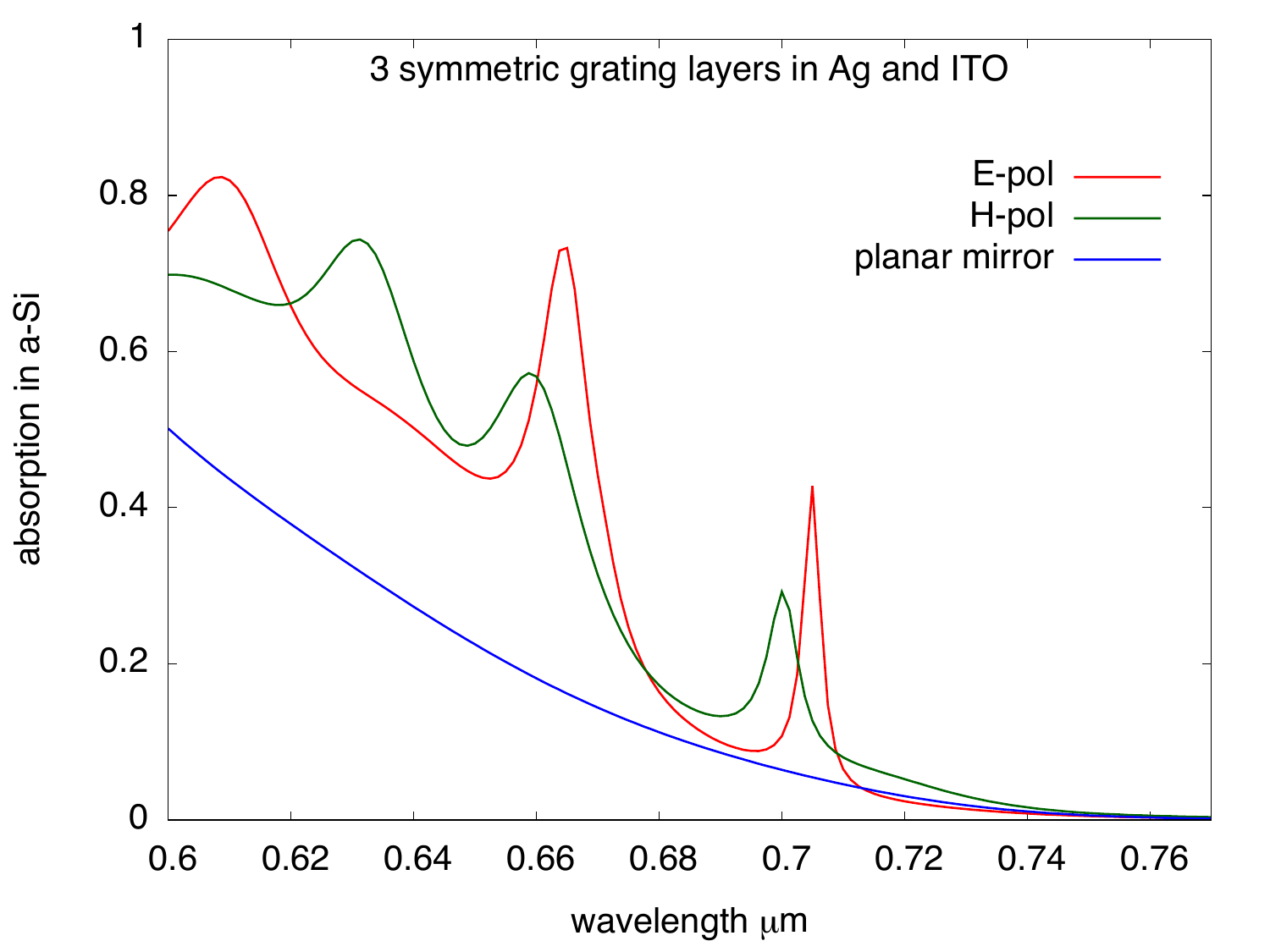}\includegraphics[clip,width=4cm]{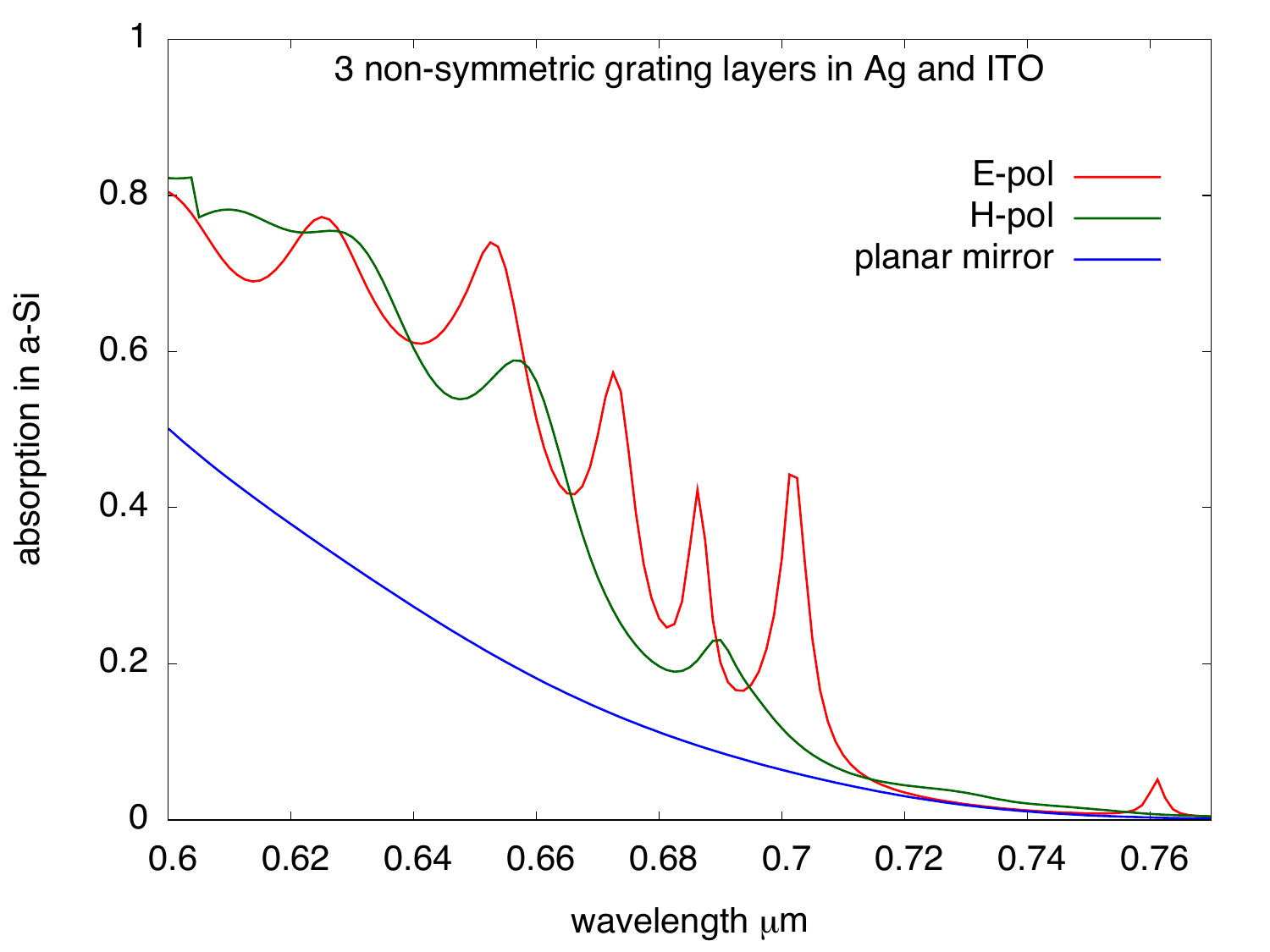}\includegraphics[clip,width=4cm]{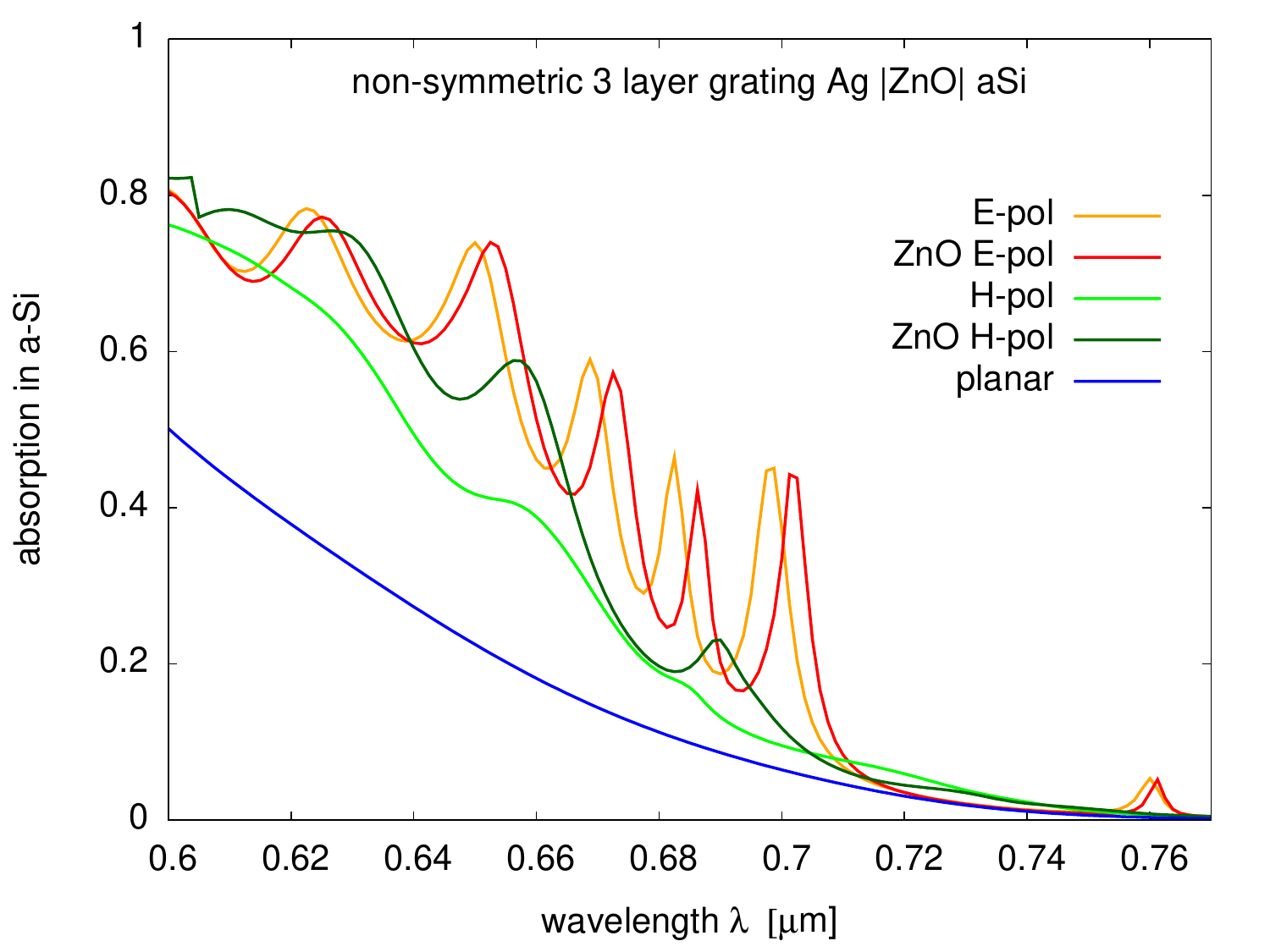}

\protect\caption{Absorption spectra of symmetric structure (left panel) and non-symmetric
structure (middle panel), both with 11nm ZnO lining, see Figs. \ref{fig:Improvement-of-integrated},\ref{fig:Non-symmetric-grating-structure}
and unlined vs. ZnO-lined non-symmetric structure (right panel).\label{fig:Absorptionspectra-symm-nonsymm}}

\end{figure}

\subsection{Absorption By Specific Modes}

The question which modes contribute most to the absorption in a specific
layer can be answered as well. To that end, the following method has
been used: First, the full boundary value problem needs to be solved,
together with the calculation of the local absorption in the chosen
layer by calculating the surface integral of the Poynting vector,
as shown in section \ref{sub:Local-Absorption}. Then, the calculation
is to be repeated, while omitting the $k^{\textrm{th}}$ eigenmode
of the chosen layer. The contribution of the $k^{\textrm{th}}$ eigenmode
to the Poynting vector becomes $\vec{S}-\vec{S_{k}},$ where $\vec{S_{k}}$
can be evaluated as $\vec{S_{k}}=\left(\vec{E}-\delta\vec{E}_{k}\right)\times\left(\vec{H}-\delta\vec{H}_{k}\right).$
Here, the $\vec{E}$ and $\vec{H}$ Fields can be evaluated as shown
in\ref{sub:Local-Absorption}, and $\delta\vec{E}_{k}$ and $\delta\vec{H}_{k}$
are the electromagnetic fields calculated by not including the $k^{\textrm{th}}$
eigenmode. 

Here, we examine the absorption within the homogeneous amorphous silicon
layer, in which the modes are just the plane wave solutions appropriate
to the dielectric properties of a-Si. The structure is the asymmetric
grating with 11nm ZnO lining illustrated in the middle panel of Fig.
\ref{fig:Absorptionspectra-symm-nonsymm} and with the structure defined
in Fig. \ref{fig:Improvement-of-integrated}.

\begin{figure}
\includegraphics[width=5.5cm]{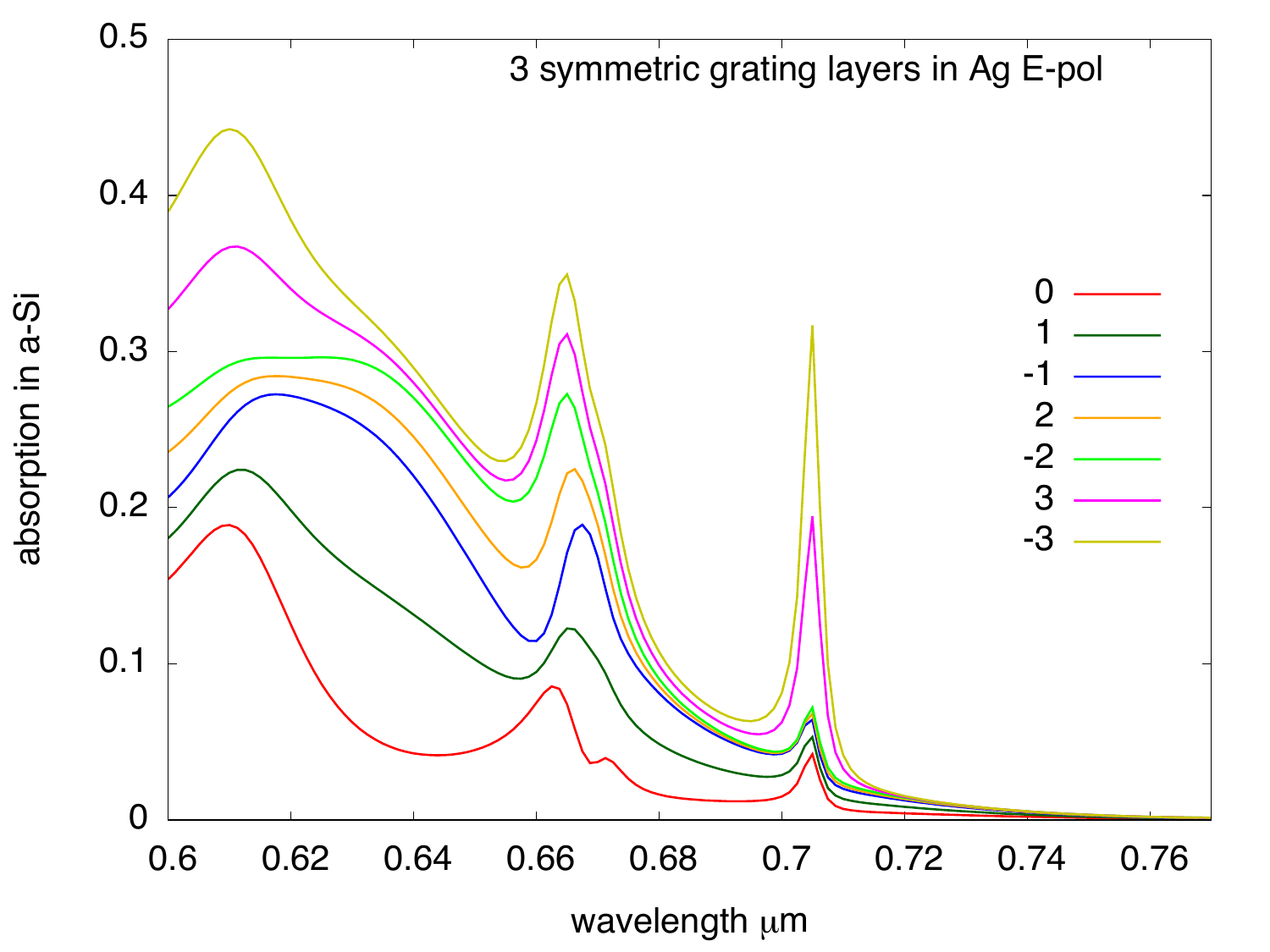}\includegraphics[clip,width=5.5cm]{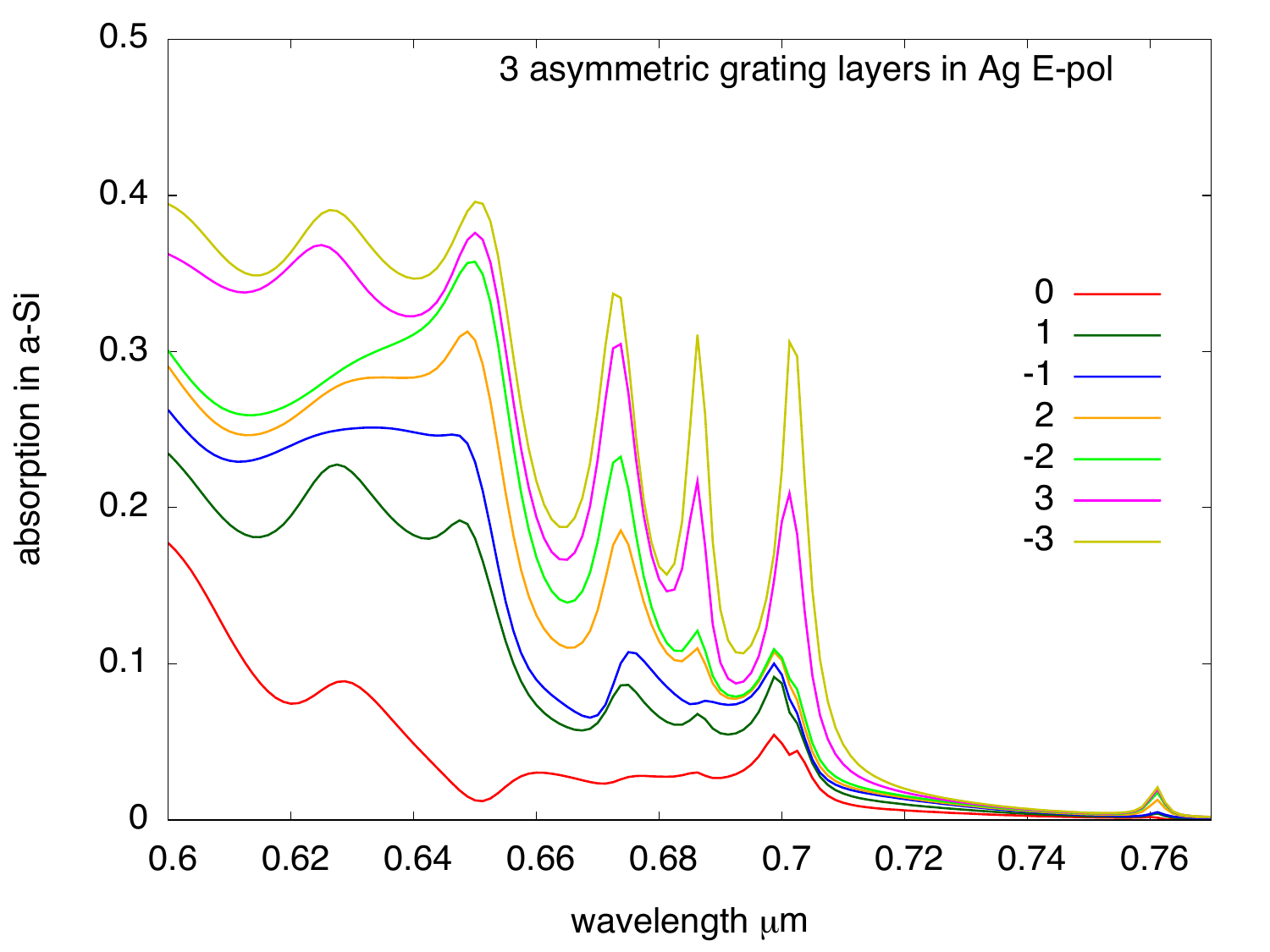}\protect\caption{Modal contribution to the absorption for a symmetric (left) and non-symmetric
(right) 3-step pseudo-sine in E polarization. Structure as in Figs.\ref{fig:Improvement-of-integrated}
and \ref{fig:Non-symmetric-grating-structure}. \label{fig:Modal-contribution-Epol-Sym} }
\end{figure}

\begin{figure}
\includegraphics[clip,width=5.5cm]{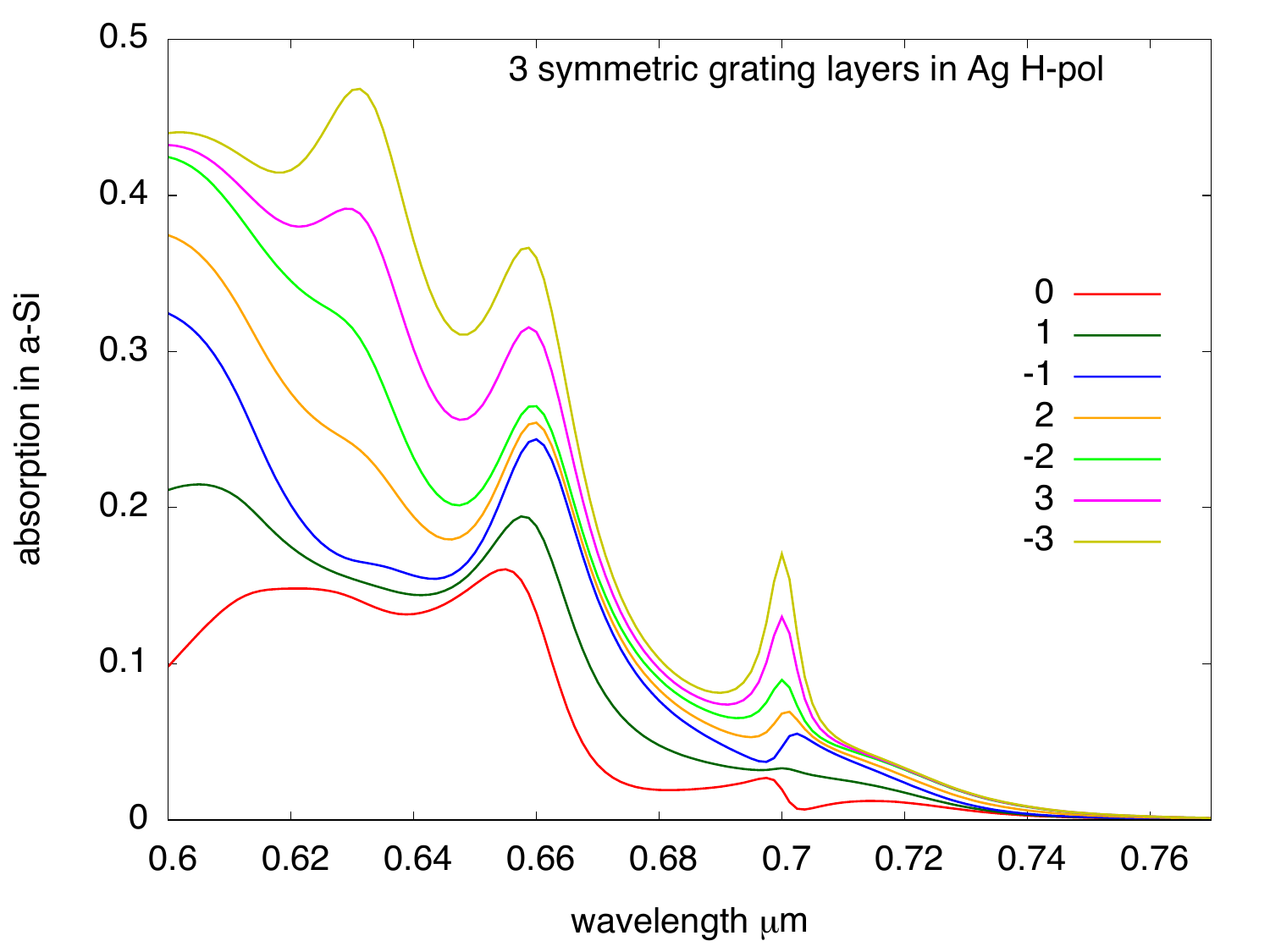}\includegraphics[clip,width=5.5cm]{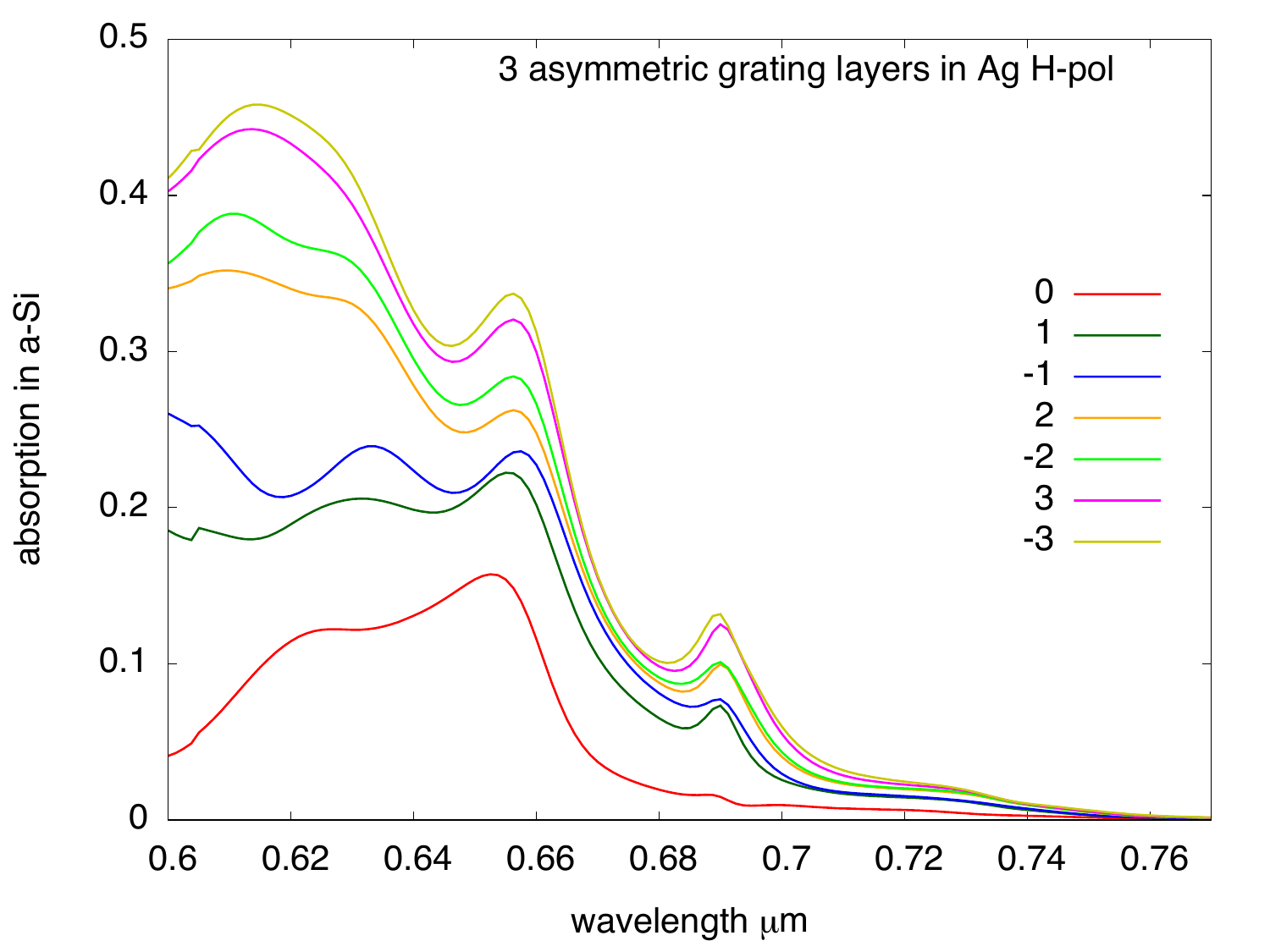}\protect\caption{Modal contribution to the absorption for a symmetric (left) and non-symmetric
(right) 3-step pseudo-sine in H polarization. Same structures as in
Fig. \ref{fig:Modal-contribution-Epol-Sym}.\label{fig:Modal-contribution-Hpol-Sym}}
\end{figure}

In the figures \ref{fig:Modal-contribution-Epol-Sym}-\ref{fig:Modal-contribution-Hpol-Sym}
the lowest of these modal contributions are shown for the case of
a homogeneous a-Si layer on top of a 3-step pseudo-sine grating. Given
that a homogeneous layer is analyzed, the eigenmodes are plane waves
and numbered according to their Fourier order. In the figures shown,
the modal contributions are consecutively added and plotted on top
of each other. These modes account for 94-99\% of all absorption within
the layer such that the curve corresponding to the mode of order -3
becomes indistinguishable with the curve of the full boundary value
problem, hence the total is not shown. Despite the fact that for all
cases shown 7 modes suffice to calculate most of the contributions
to absorption within the layer, the higher order modes are needed
to calculate the boundary value problem accurately, and for the following
calculations, 31 numerically exact modes in each layer were used.

Figures \ref{fig:Modal-contribution-Epol-Sym} and \ref{fig:Modal-contribution-Hpol-Sym}
show the modal contributions for E-polarization and H-polarization
respectively on a symmetric and non-symmetric grating, and it can
be seen that modes of equal number contribute an equal amount for
both polarization. Furthermore, it is worth noting that throughout
the spectrum, many modes contribute significantly to the absorption,
whereas the resonance above $0.7\,\textrm{\ensuremath{\mu}m}$ shows
a more dominant contribution of the $\pm3$ mode for the absorption.
In contrast, a similar resonance exists for H-polarization, where
multiple modes contribute to the absorption.

Figures \ref{fig:Modal-contribution-Epol-Sym} (right panel) and \ref{fig:Modal-contribution-Hpol-Sym}(right
panel) show the modal contributions for E-polarization and H-polarization
respectively on an non-symmetric grating with the same grating parameters
as above, but where all grating layers are shifted by $0.115\times\Lambda$
with respect to neighboring layers, such that the resulting grating
shows a significant asymmetry. Correspondingly, the modes of equal
number do not contribute equal amounts to the absorption. For E-polarization
the spectrum shows multiple resonances, where some modes contribute
visibly more than others. In particular, the resonances above $0.68\,\textrm{\ensuremath{\mu}m}$
and above $0.7\,\textrm{\ensuremath{\mu}m}$ seem to have a significant
contribution from the modes of order $\pm3$, however throughout the
spectrum the other modes contribute visibly. For H-polarization, there
are less visible resonances, and around $0.66\,\textrm{\ensuremath{\mu}m}$
the background absorption begins to drop. There, the relative contribution
of the mode of order $\text{+1}$ becomes the dominant contribution,
until about $0.71\,\textrm{\ensuremath{\mu}m}$, where the total absorption
becomes insignificant altogether.

\section{Discussion}

In this paper, we have first developed an accurate and efficient method
to calculate the material specific absorption with our numerically
exact modal method. The accuracy of the solution is determined solely
by the truncation order N, once the eigenvalues and eigenfunctions
are ordered according to their relevance. 

The criteria for the validity of calculations are convergence as the
truncation order is increased, and we calculate the energy conservation
and convergence with respect to the approximation of a given smooth
surface profile, for which stair-case approximations are made.

For H-polarization, we have observed large parasitic absorption in
metallic gratings in the spectral range in which the absorptivity
of the semiconductor is weak. We trace this back to large currents
close to the interfaces between metal and semiconductor. Indeed, we
have found two-fold nearly degenerate eigenvalues of the Helmholtz
equation for such gratings in an extended spectral range, and even
numerically exactly degenerate for a specific wavelength. At normal
incidence, the associated eigenfunctions are either symmetric or antisymmetric.
They are localized at the position of the metal-semiconductor interface
and decay exponentially away from it. Thus their degeneracy follows
as the tunnel splitting will decay also exponentially as a function
of the separation of the two interfaces. 

Based on the presented findings, we attribute a significant part of
these difficulties to interface plasmons and all modes that are responsible
for large currents at and close to interfaces between metals and semiconductors.
These currents are the source of the large parasitic absorption in
even the metals with the smallest imaginary part of the dielectric
constant, e.g. silver. For this reason, the ideal limit of absorption-free
metals is so unrealistic that it is of no interest for photovoltaics.

We briefly summarize the most important findings of this work: Light-trapping
can be improved by optimizing grating depth and period, by introducing
an asymmetry which allows a coupling between modes of even and odd
parity, effectively doubling the mode space. Asymmetry is found to
be more effective for E-polarization. For some grating structures,
it is found that the absorption in the semiconductor can increase
significantly when its thickness is somewhat reduced. The analysis
of the contribution to the absorption in the semiconductor of individual
modes shows that a number of modes is of nearly equal importance even
near and at resonance peaks in the spectrum. Finally, we find that
the parasitic absorption in a metal-semiconductor grating can be reduced
by insertion of a rather thin layer of a dielectric, in our case of
the order of zinc-oxide of 10nm width.

In view of all these results, it is unclear to us, whether we can
expect much better light-trapping performance from two-dimensional
metal gratings. In that case, all polarizations will suffer from the
problem of large interfacial currents and consequently enhanced parasitic
absorption as we observe in one-dimensional metal gratings in H-polarization.

On the other hand, two-dimensional gratings between dielectric and
semiconducting materials may well turn out superior. Modal calculations
allowing a fairly exhaustive investigation of such structures will
help to answer this question.

\section{Acknowledgments}

We acknowledge the financial support of this work by the Swiss Federal
Office of Energy, and by the Paul Scherrer Institute. We also would
like to thank F.J. Haug and H.P.Herzig for helpful discussions and
advice, and E. Barakat and G. Smolik for reference calculations.

\section{Appendix}

\section*{\textcolor{black}{Dispersion Fits}}

\textcolor{black}{For polycarbonate, a refractive index of $n=1.58$
was used in the calculations.}

\textcolor{black}{Whenever noted in the grating calculations, a fit
function was used to parametrize measured data. In particular, for
the amorphous silicon, the dispersion data was supplied by Franz-Josef
Haug, and the following fit functions were used for parametrization:}

\textcolor{black}{
\begin{eqnarray*}
Re[n] & = & exp(f(\lambda))\\
Im[n] & = & exp(g(\lambda))
\end{eqnarray*}
}

where $f(\lambda)$ and $\text{g(\ensuremath{\lambda}})$ are fitted
by polynomials

\textcolor{black}{
\begin{alignat*}{1}
f\left(\lambda\right) & =f_{0}+\lambda(f_{1}+\lambda(f_{2}+\lambda\, f_{3}))\\
g(\lambda) & =g0+\lambda(g_{1}+\lambda(g_{2}+\lambda(g_{3}+\lambda\, g_{4}))),
\end{alignat*}
}

\begin{table}
\begin{tabular}{|c|r@{\extracolsep{0pt}.}l|r@{\extracolsep{0pt}.}l|r@{\extracolsep{0pt}.}l|r@{\extracolsep{0pt}.}l|r@{\extracolsep{0pt}.}l|}
\hline 
 & \multicolumn{2}{c|}{{\footnotesize{}a-Si}} & \multicolumn{2}{c|}{{\footnotesize{}ZnO}} & \multicolumn{2}{c|}{{\footnotesize{}ITO}} & \multicolumn{2}{c|}{{\footnotesize{}Ag}} & \multicolumn{2}{c|}{{\footnotesize{}Al}}\tabularnewline
\hline 
\hline 
{\footnotesize{}f0} & {\footnotesize{}0}&{\footnotesize{}40404338} & {\footnotesize{}1}&{\footnotesize{}70771866} & {\footnotesize{}1}&{\footnotesize{}21219668} & {\footnotesize{}2}&{\footnotesize{}29527609} & {\footnotesize{}-2}&{\footnotesize{}26656817}\tabularnewline
\hline 
{\footnotesize{}f1} & {\footnotesize{}7}&{\footnotesize{}66872671} & {\footnotesize{}-4}&{\footnotesize{}37318385} & {\footnotesize{}-2}&{\footnotesize{}14526761} & {\footnotesize{}-19}&{\footnotesize{}04662349} & {\footnotesize{}3}&{\footnotesize{}19251777}\tabularnewline
\hline 
{\footnotesize{}f2} & {\footnotesize{}-15}&{\footnotesize{}06785446} & {\footnotesize{}6}&{\footnotesize{}12742066} & {\footnotesize{}2}&{\footnotesize{}91527508} & {\footnotesize{}26}&{\footnotesize{}97072796} & {\footnotesize{}2}&{\footnotesize{}33399365}\tabularnewline
\hline 
{\footnotesize{}f3} & {\footnotesize{}8}&{\footnotesize{}73093040} & {\footnotesize{}-2}&{\footnotesize{}98521339} & {\footnotesize{}-1}&{\footnotesize{}78130071} & {\footnotesize{}-12}&{\footnotesize{}13112197} & {\footnotesize{}-1}&{\footnotesize{}43428367}\tabularnewline
\hline 
{\footnotesize{}g0} & {\footnotesize{}19}&{\footnotesize{}65451137} & {\footnotesize{}184}&{\footnotesize{}89403859} & {\footnotesize{}-10}&{\footnotesize{}55809870} & {\footnotesize{}-3}&{\footnotesize{}25349091} & {\footnotesize{}-3}&{\footnotesize{}09108230}\tabularnewline
\hline 
{\footnotesize{}g1} & {\footnotesize{}-107}&{\footnotesize{}24524864} & {\footnotesize{}-1'331}&{\footnotesize{}94129402} & {\footnotesize{}18}&{\footnotesize{}52031840} & {\footnotesize{}16}&{\footnotesize{}14691273} & {\footnotesize{}29}&{\footnotesize{}11819344}\tabularnewline
\hline 
{\footnotesize{}g2} & {\footnotesize{}210}&{\footnotesize{}41585170} & {\footnotesize{}3'398}&{\footnotesize{}09758613} & {\footnotesize{}-17}&{\footnotesize{}55613183} & {\footnotesize{}-19}&{\footnotesize{}20517240} & {\footnotesize{}-70}&{\footnotesize{}78995115}\tabularnewline
\hline 
{\footnotesize{}g3} & {\footnotesize{}-153}&{\footnotesize{}50212937} & {\footnotesize{}-3'794}&{\footnotesize{}48410470} & {\footnotesize{}7}&{\footnotesize{}49002208} & {\footnotesize{}8}&{\footnotesize{}36374561} & {\footnotesize{}82}&{\footnotesize{}73715473}\tabularnewline
\hline 
{\footnotesize{}g4} & {\footnotesize{}0}&{\footnotesize{}00000000} & {\footnotesize{}1'570}&{\footnotesize{}98004140} & {\footnotesize{}0}&{\footnotesize{}00000000} & {\footnotesize{}0}&{\footnotesize{}00000000} & {\footnotesize{}-36}&{\footnotesize{}92328757}\tabularnewline
\hline 
\end{tabular}\protect\caption{Dispersion data for a-Si, Ag, Al, ZnO and ITO}

\end{table}

\textcolor{black}{In Table 1, we list the fitting parameters which
assume that the wavelength is measured in micrometers for all materials,
that are used in the present study of light-trapping performance. }

\section*{\textcolor{black}{Anti-Reflective Coating}}

\textcolor{black}{For many calculations in this thesis, optimum limits
for the absorption within the semiconductor layer are a goal. For
a setting with light incident from the outside, reflection at the
top surface must be suppressed to the extent possible. To do so, the
following artificial antireflective coating was used in some calculations.
This idealized cover almost completely eliminates reflection of the
incident light at the top interface of the structure by introducing
a homogeneous layer on top of the structure, which is made up of an
artificial material AR{*} whose thickness is adjusted as a function
of wavelength. This is clearly unphysical, but permits us to examine
specifically light-trapping efficiency.}

\textcolor{black}{This material AR{*} is supposed to have a wavelength
dependent refractive index $n^{*}(\lambda)=\sqrt{\Re\, n_{a-Si}(\lambda)}$,
where $n_{a-Si}$ is the refractive index of the layer below, and
the $\Re$ symbol stands for the real part. If the antireflection
coating is taken to have a thickness corresponding to $\frac{\lambda}{4n^{*}(\lambda)}$
, the reflections can be suppressed except at the ultraviolet and
blue part of the spectrum, where amorphous silicon absorbs very efficiently.
While this idealized device is unphysical, there are realizations
of antireflective structures which lead to similarly low reflexion
losses \citep{Heine:1995qw,Heine:1996zv}. The broadband antireflection
structure described therein consists of a rectangular grating with
an additional coating with a lower refractive index than the grating
below. We did not want to introduce the complications resulting from
having gratings with different period for antireflection and light-trapping.
That such combinations will work in practice is certain. But it falls
out of the scope of the present study.}

\section*{Table of step height and positions for mimicking sine-grating by
stair-case }

\begin{table}
\begin{tabular}{|c|c|c|c|c|c|c|}
\hline 
layer $i$ &  & 1 & 2 & 3 & 4 & 5\tabularnewline
\hline 
$P^{(1)}$ & $h$ $[A]$ & 1 &  &  &  & \tabularnewline
\hline 
 & $h(i)$ $[A]$ & $0.5$ &  &  &  & \tabularnewline
\cline{2-7} 
$P^{\left(2\right)}$ & $x_{1}$ $[\Lambda]$ & $0.340845$ &  &  &  & \tabularnewline
\hline 
 & $h(i)$ $[A]$ & $0.30981$ & $0.38037$ &  &  & \tabularnewline
\cline{2-7} 
$P^{\left(3\right)}$ & $x_{1}$ $[\Lambda]$ & $0.377796$ & $0.25$ &  &  & \tabularnewline
\hline 
 & $h(i)$ $[A]$ & $0.16415$ & $0.21939$ & $0.23292$ &  & \tabularnewline
\cline{2-7} 
$P^{\left(5\right)}$ & $x_{1}$ $[\Lambda]$ & $0.424452$ & $0.337468$ & $0.25$ &  & \tabularnewline
\hline 
 & $h(i)$ $[A]$ & $0.10676$ & $0.14646$ & $0.16289$ & $0.16777$ & \tabularnewline
\cline{2-7} 
$P^{\left(7\right)}$ & $x_{1}$ $[\Lambda]$ & $0.429895$ & $0.361367$ & $0.303917$ & $0.25$ & \tabularnewline
\hline 
 & $h(i)$ $[A]$ & $0.077125$ & $0.10711$ & $0.12144$ & $0.12879$ & $0.13109$\tabularnewline
\cline{2-7} 
$P^{\left(9\right)}$ & $x_{1}$ $[\Lambda]$ & $0.440601$ & $0.383026$ & $0.335513$ & $0.291968$ & $0.25$\tabularnewline
\hline 
\end{tabular}

\protect\caption{The optimised step position $x(i)$ and heigths $h(i)$ of pseudo\_sine
structures for N=1,2,3,5,7,9 steps.\protect \\
Let $M=Int[N/2]$, $h(N+1-i)=h(i)$, $x_{1}(N+1-i)=\frac{1}{2}-x_{1}(i),i=1,2,...,M$
and $x_{2}(i)=1-x_{1}(i)$: then material I is in layer $i$ inside
the interval ($x_{1}(i),x_{2}(i))$ and material II is in the remaining
part of the interval (0,1).\label{tab:The-optimised-step}}
\end{table}

\bibliographystyle{apsrev4-1}
\bibliography{reduced}

\end{document}